\documentclass[aps,prd,preprint,superscriptaddress,tightenlines,nofootinbib]{revtex4}

\usepackage{graphicx}% Include figure files
\usepackage{dcolumn}% Align table columns on decimal point
\usepackage{bm}% bold math

\begin{document}

\preprint{CLNS 06/1980,  CLEO 06-20}   % For conference papers
\title{$\chi_{cJ}$ Decays to $h^+h^-h^0$}

%-------- INSERT HERE ------------
% Your author list goes here  REMOVE EVERYTHING to END INSERT and
% replace with your authorlist (ask cleoac).
%\input ./clns-06-1980.tex

\author{S.~B.~Athar}
\author{R.~Patel}
\author{V.~Potlia}
\author{J.~Yelton}
\affiliation{University of Florida, Gainesville, Florida 32611}
\author{P.~Rubin}
\affiliation{George Mason University, Fairfax, Virginia 22030}
\author{C.~Cawlfield}
\author{B.~I.~Eisenstein}
\author{I.~Karliner}
\author{D.~Kim}
\author{N.~Lowrey}
\author{P.~Naik}
\author{M.~Selen}
\author{E.~J.~White}
\author{J.~Wiss}
\affiliation{University of Illinois, Urbana-Champaign, Illinois 61801}
% ...
%\author{(CLEO Collaboration)} %FOR PRD_SPECIAL_CHANGEME
\collaboration{CLEO Collaboration} %FOR PRL,CLNS
\noaffiliation

%-------- END INSERT ------------

%please hard code the date when you have a final draft and submit to CLEOAC
%\date{\today}
\date{February 9, 2006}

\begin{abstract} 
Using a sample of $3 \times 10^6$ $\psi(2S)$ decays recorded by the CLEO detector,
we study three-body decays of the $\chi_{c0}$, $\chi_{c1}$, and
$\chi_{c2}$ produced in radiative decays of the $\psi(2S)$.  
We consider the final states $\pi^+\pi^-\eta$, $K^+K^-\eta$, $p{\bar p}\eta$,
$\pi^+\pi^-\eta^\prime$, $K^+K^-\pi^0$, $p{\bar p}\pi^0$, $\pi^+K^-K^0_{\rm S}$, and
$K^+{\bar p}\Lambda$, measuring branching fractions or placing upper limits.
For $\chi_{c1} \to \pi^+\pi^-\eta$,  $K^+K^-\pi^0$,
and $\pi^+K^-K^0_{\rm S}$ our observed samples are large enough 
to indicate the largest contributions to the substructure.
\end{abstract}

\maketitle

%%%======================================================================
\section{Introduction}

	Decays of the $\chi_{c0}$, $\chi_{c1}$, and $\chi_{c2}$ states 
are not as well studied 
experimentally and theoretically as those of other charmonium states.  
These states have charge-conjugation eigenvalue $C = +1$, in contrast to
the better-studied $C = -1$ states $J/\psi$ and $\psi(2S)$.  Their decay
products thus will differ from those of $J/\psi$ and $\psi(2S)$, and may
provide complementary information on states containing light quarks
and gluons.
It is possible that the color-octet mechanism,
$c \bar c g \to 2(q \bar q)$, could have large effects on the observed
decay pattern of the $\chi_{cJ}$ states~\cite{quarkoniumreview}.
Assuming that $\chi_{cJ}$ are the $^3P_J$ $c\bar{c}$ bound states
one would expect that $\chi_{c0}$ and $\chi_{c2}$ with $J^{PC}$ quantum numbers
$0^{++}$ and $2^{++}$ decay to light quarks via two gluons~\cite{Zhao}.
Measurement of any possible $\chi_{cJ}$ hadronic decays provides
valuable information on possible glueball dynamics.
Thus knowledge of any hadronic decay channels for these
states is valuable.

	CLEO has gathered a large sample of $e^+e^- \to \psi(2S)$ events,
which leads to copious production of the $\chi_{cJ}$ states in
radiative decays of the $\psi(2S)$.
The $\psi(2S)$ branching fractions
have been recently measured \cite{chicbf} with high precision:
\begin{equation}
\label{eqn:br_chic0}
    {\mathcal B}(\psi(2S)\to\gamma \chi_{c0}) = (9.22\pm 0.11\pm 0.46)\%;
\end{equation}
\begin{equation}
\label{eqn:br_chic1}
    {\mathcal B}(\psi(2S)\to\gamma \chi_{c1}) = (9.07\pm 0.11\pm 0.54)\%;
\end{equation}
\begin{equation}
\label{eqn:br_chic2}
    {\mathcal B}(\psi(2S)\to\gamma \chi_{c2}) = (9.33\pm 0.14\pm 0.61)\%.
\end{equation}
We describe
a study of selected three-body hadronic decay modes of the
$\chi_{cJ}$ to two charged and one neutral hadron.
This is not an exhaustive study of $\chi_{cJ}$ hadronic decays;
we do not even comprehensively cover all possible $h^+h^-h^0$ decays,
but simply take a first look
at the rich structure of $\chi_{cJ}$ decays in our initial
$\psi(2S)$ data sets.  A subset of these modes has
been investigated by BES~\cite{BES}.
With the CLEO III detector configuration~\cite{CLEOIII}, 
we have recorded an integrated luminosity of 
2.57~pb$^{-1}$ and the number of $\psi(2S)$ events
is $1.56 \times 10^6$.
With the CLEO-c detector configuration~\cite{CLEOc} we have recorded
2.89~pb$^{-1}$, and the number
of events is $1.52 \times 10^6$. The apparent mis-match of luminosities 
and event totals is due to different beam energy spreads for the two data sets.

%%%======================================================================
\section{Measurement of the branching fractions}

	Our basic technique is an exclusive whole-event analysis searching
for $\psi(2S)\to\gamma\chi_{cJ}$ followed by a three body
decay of the $\chi_{cJ}$ to $\pi^+\pi^-\eta$, $K^+K^-\eta$, $p{\bar p}\eta$,
$\pi^+\pi^-\eta^\prime$, $K^+K^-\pi^0$, $p{\bar p}\pi^0$, $\pi^+K^-K^0_{\rm S}$, or
$K^+{\bar p}\Lambda$.  A photon
candidate is combined with three hadrons and their 4-momentum sum constrained
to the known total beam energy and the initial momentum caused by the two 
beams crossing at a small angle taking into account the measured errors on
the reconstructed charged tracks, neutral hadron, and transition
photon.  We cut on the $\chi^2$ of this fit, which has four degrees of freedom,
as it strongly discriminates between background and signal.  
For most modes we select events with an event 4-momentum fit 
$\chi^2$ less than 25, but
background from $\psi(2S)\to J/\psi\pi^0\pi^0$ followed by charged 
two-body decays of the $J/\psi$, with one of the $\pi^0$
decay photons lost, fakes $\psi(2S)\to\gamma\chi_{cJ}\to \gamma p{\bar p}\pi^0$.
For this mode the cut on $\chi^2$ is tightened to 12.  The
efficiency of this cut is $\approx$ 95\% for all modes except
$p{\bar p}\pi^0$, where it is $\approx$ 80\%.  
%Little background survives.

Efficiencies
and backgrounds are studied in a GEANT-based simulation~\cite{cleog} of
the detector response to $e^+e^- \to \psi(2S)$ events.
Our simulated sample is roughly ten times our data sample.
The radiated photon is generated according to an
angular distribution of $ 1 + \lambda \cos^2\theta$,
where $\theta$ is the radiated photon angle 
relative to the positron beam axis. 
An E1 transition, as expected for $\psi(2S) \to \gamma\chi_{cJ}$,
implies $\lambda=1,-1/3,+1/13$ for $J=0,1,2$ particles. 
The efficiencies we quote use this simulation,
and differ from efficiencies using $\lambda=0$ by up to a few percent.
%The differences of efficiencies 
%due to various $\theta$ distributions are negligible
%as we accept transition photons with energies down to our detection limit
%of 30 MeV.

	Photon candidates are selected by their 
energy depositions in the CsI crystal calorimeter.
They have a transverse shape consistent with that expected for an electromagnetic
shower without a charged track pointing toward it.  
They are required to have an energy of at least 30 MeV.  
Photon candidates that are used to make
neutral particles further must
have an energy of more than 50 MeV if they are not in the barrel, $|\cos\theta_\gamma| > 0.82$,
of our calorimeter. The $\pi^0 \to \gamma\gamma$ and $\eta \to \gamma\gamma$
candidates are formed from two-photon candidates that are kinematically fit to
the known resonance masses
using the event vertex position, determined using charged tracks constrained to
the beam spot.  We select events with a $\chi^2$ from the kinematic mass
fit with one degree of freedom of less than 10.
Transition photon candidates are vetoed if they form a
$\pi^0$ or $\eta$ candidate when paired with a second photon candidate.

We also reconstruct the $\eta \to \pi^+\pi^-\pi^0$ mode combining
two charged pions with a $\pi^0 \to \gamma\gamma$ candidate, increasing the 
number of $\eta$ candidates  
by about 25\%.  The same sort of kinematic mass fit
as used for $\pi^0$'s and $\eta \to \gamma\gamma$
is applied to this mode, and again we select those giving a $\chi^2$
of less than 10.
Similarly we combine the mass-constrained $\eta$ candidates together with two charged pions to 
make $\eta^{\prime}$ candidates, mass-constrain them, and select those with $\chi^2 < 10$. 
In addition, we include the decay mode $\eta^{\prime} \to \gamma \rho$. 
Here the background is potentially high because of the large 
number of noise photons, so we require 
$E_{\rm photon} > 200$~MeV. In addition, we require the $\pi^+\pi^-$ mass
to be within 100~MeV/c$^2$ of the mean $\rho$ mass.  

Charged tracks satisfy standard requirements \cite{HadronicBF} that they
be of good fit quality. 
Those coming from the origin must have an impact parameter
with respect to the beam spot less than the
greater of $(5.0-3.8\cdot p)$~mm and 1.2~mm, where $p$ is the measured
track momentum in GeV/c.
The $K^0_S \to \pi^+\pi^-$ and $\Lambda \to p\pi^-$ candidates
are formed from good-quality tracks that are constrained to come
from a common vertex.
The $K^0_S$ flight path is required to be greater than 5~mm and the 
$\Lambda$ flight path greater than 3~mm.  The mass cut around the $K^0_S$ mass is 
$\pm 10$~MeV/c$^2$, and around the $\Lambda$ mass $\pm 5$~MeV/c$^2$,
both about three times the resolution.
Events with only the exact number of selected tracks
are accepted.
This selection is very efficient, $>$99.9\%, for events
passing all other requirements.

	Pions are required to have specific ionization,
$dE/dx$, in the main drift chamber within four standard
deviations of the expected value for
a real pion at the measured momentum. 
For kaons and protons, a combined $dE/dx$ and RICH
(ring imaging Cherenkov counter) 
likelihood is formed and kaons are required
to be more kaon-like than pion- or proton-like, 
and similarly for protons.
Cross feed between hadron species is negligible after all
other requirements. 

In modes comprising only two charged particles, 
there are some extra cuts to eliminate QED background which
produce charged leptons in the final state.
Events are rejected if the sum over all the charged
tracks produces a penetration into the muon system of more
than five nuclear interaction lengths.
Events are rejected if any track has $0.92<E/p<1.05$ and it
has a $dE/dx$ consistent with an electron.
This latter cut is not used for $p\bar{p}$ 
modes because anti-protons tend to deposit all their
energy in the calorimeter.  These cuts are essentially 100\% efficient
for the signal,
and ensure this QED background is negligible.

The efficiencies averaged over the CLEO~III and CLEO-c
data sets for each mode including the branching fractions
$\eta \to \gamma\gamma$, $\eta \to \pi^+\pi^-\pi^0$, and $\eta^\prime \to \eta\pi^+\pi^-$~\cite{pdg}
are given in Tables~\ref{tab:BRfits0}-\ref{tab:BRfits2} for
$\chi_{c0}$,
$\chi_{c1}$, and
$\chi_{c2}$ respectively.
\begin{figure}
\begin{minipage}[t]{75mm}
\includegraphics*[width=72mm]{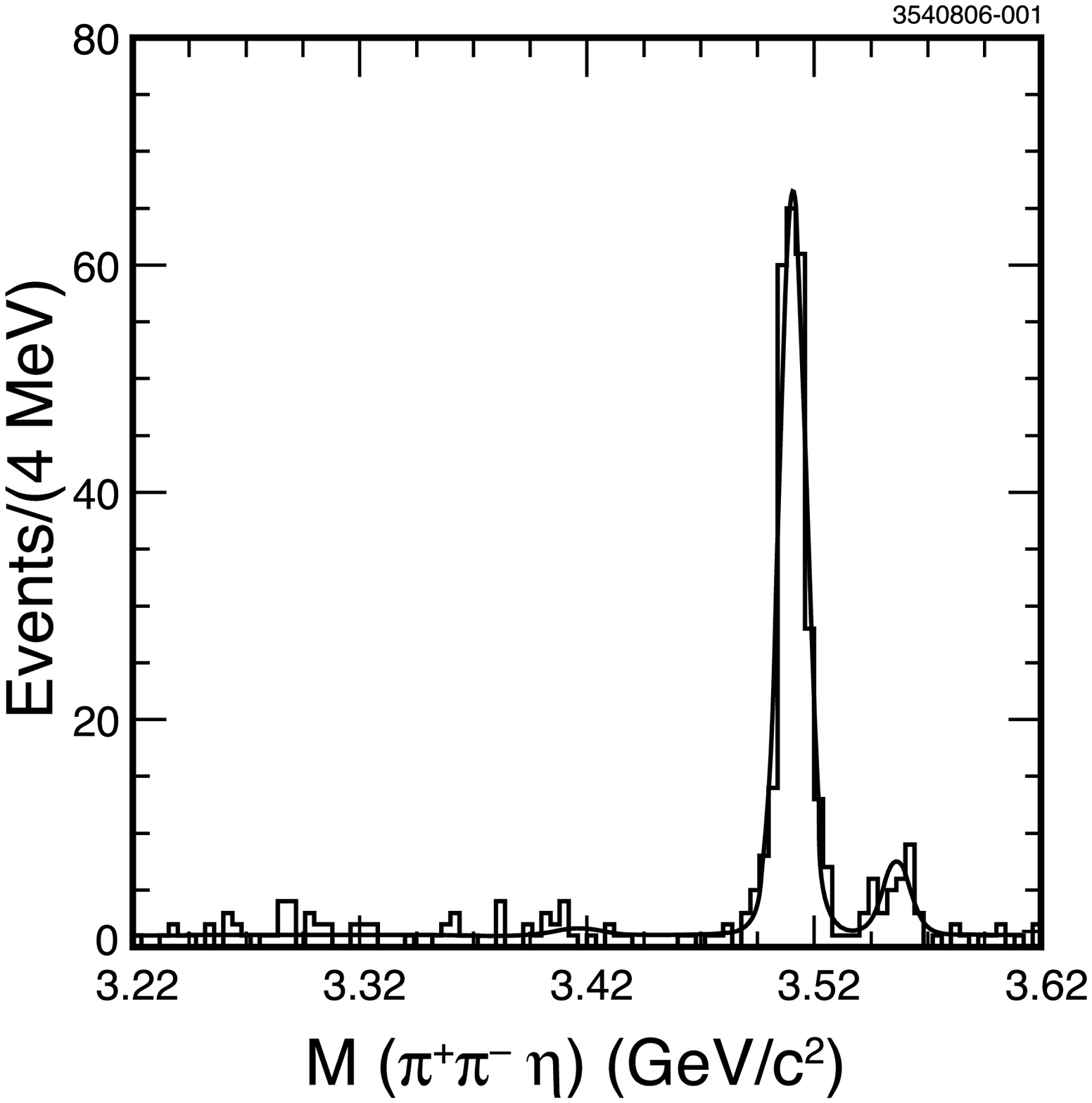}
\caption{Mass distribution for candidate $\chi_{cJ} \to \pi^+\pi^-\eta$ events.
         The displayed fit is described in the text.}
\label{fig:pipieta}
\end{minipage}
\hfill
\begin{minipage}[t]{75mm}
\includegraphics*[width=72mm]{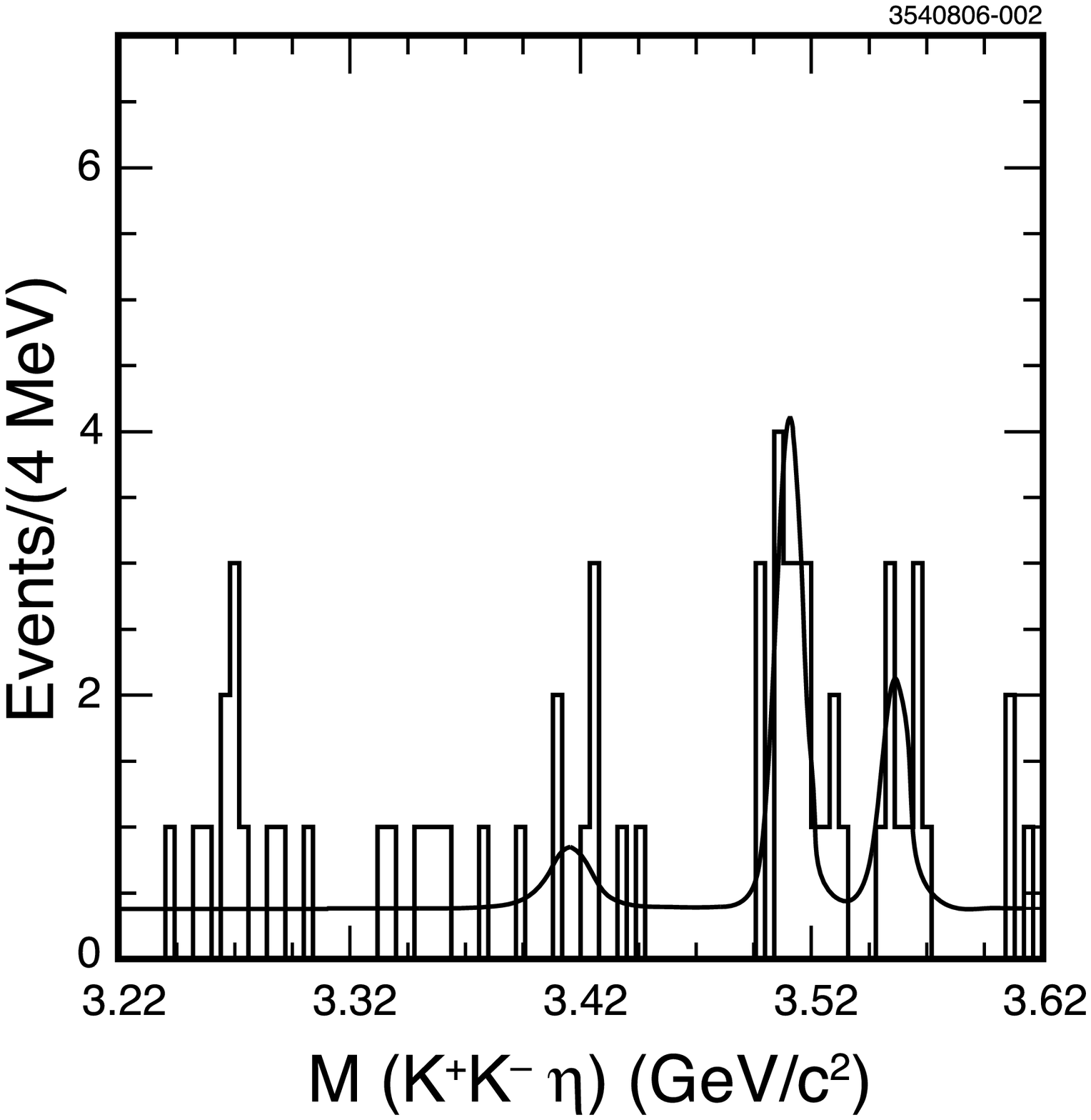}
\caption{Mass distribution for candidate $\chi_{cJ} \to K^+K^-\eta$ events.
         The displayed fit is described in the text.}
\label{fig:KKeta}
\end{minipage}
\begin{minipage}[t]{75mm}
\includegraphics*[width=72mm]{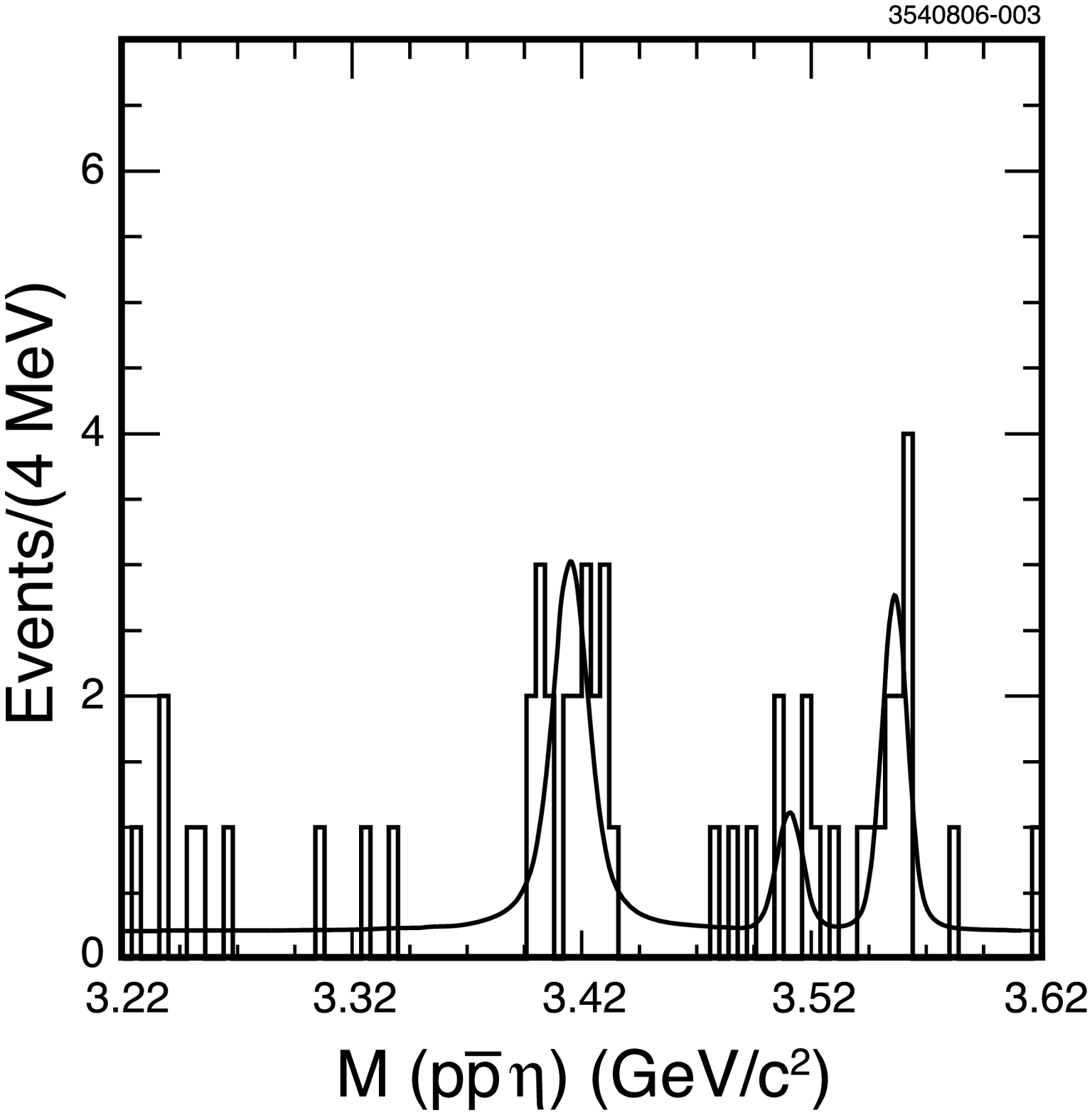}
\caption{Mass distribution for candidate $\chi_{cJ} \to p{\bar p}\eta$ events.
         The displayed fit is described in the text.}
\label{fig:ppbeta}
\end{minipage}
\hfill
\begin{minipage}[t]{75mm}
\includegraphics*[width=72mm]{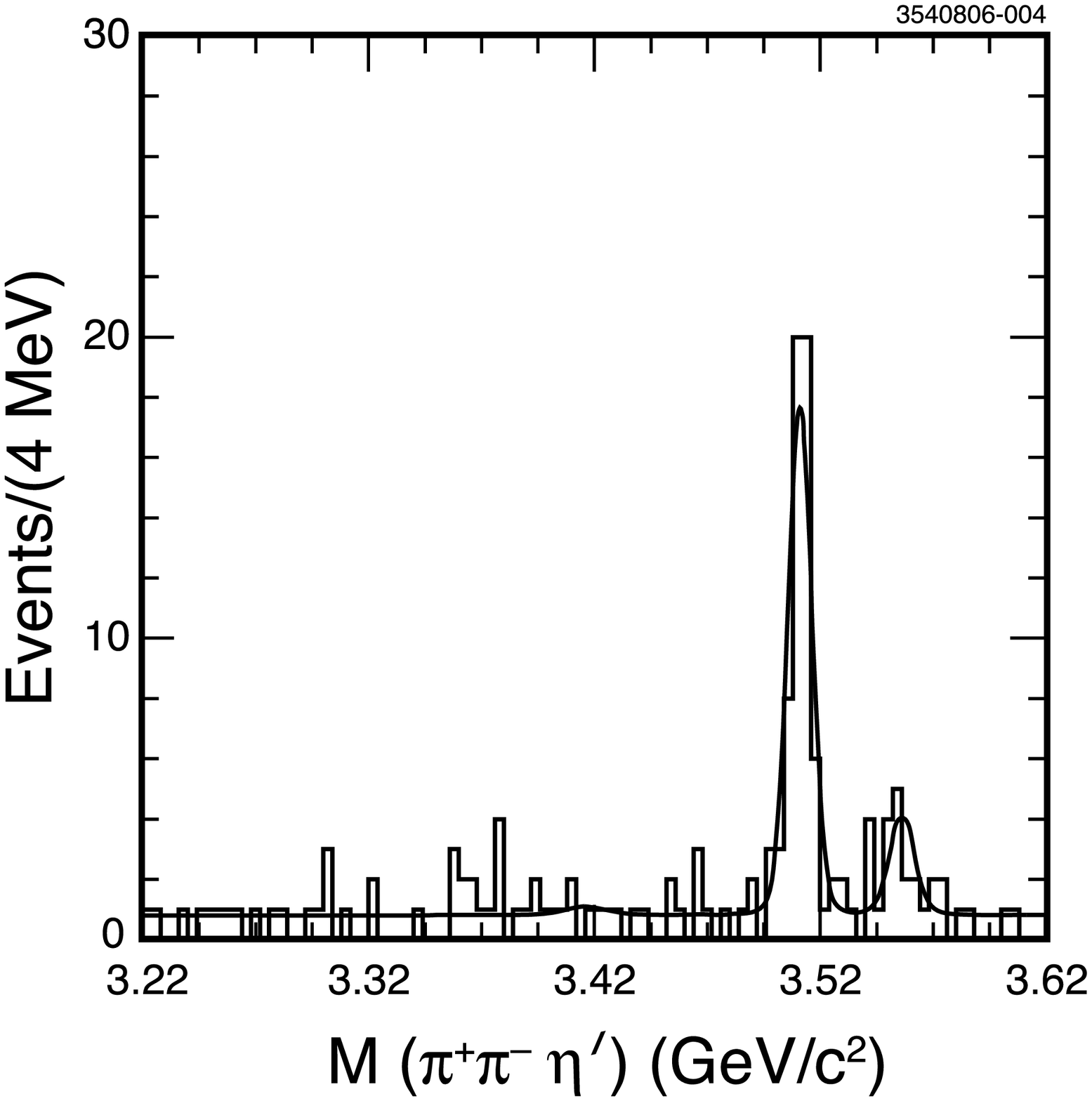}
\caption{Mass distribution for candidate $\chi_{cJ} \to \pi^+\pi^-\eta^\prime$ events.
         The displayed fit is described in the text.}
\label{fig:pipietaP}
\end{minipage}
\end{figure}
\begin{figure}
\begin{minipage}[t]{75mm}
\includegraphics*[width=72mm]{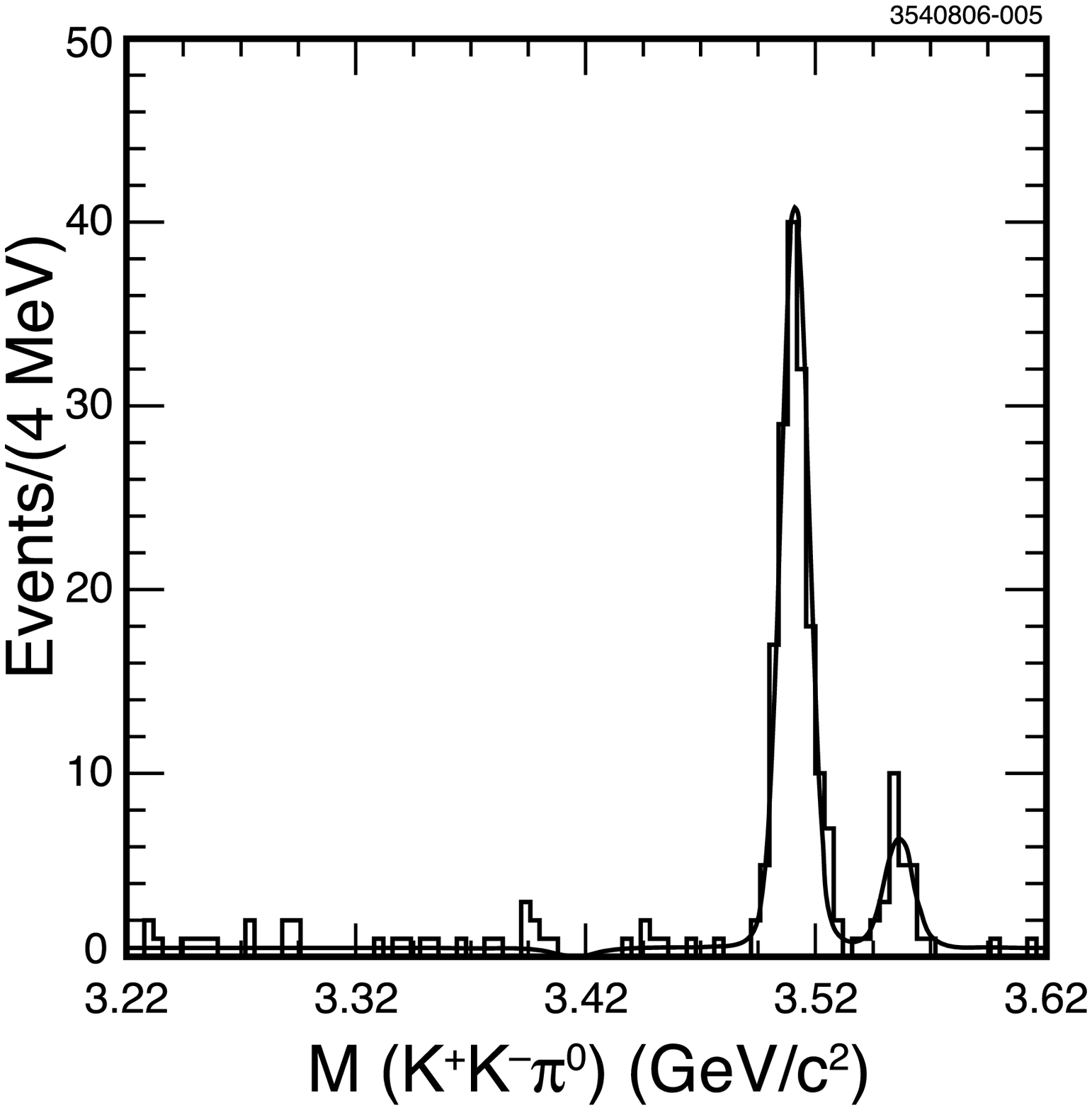}
\caption{Mass distribution for candidate $\chi_{cJ} \to K^+K^-\pi^0$ events.
         The displayed fit is described in the text.}
\label{fig:KKpi0}
\end{minipage}
\hfill
\begin{minipage}[t]{75mm}
\includegraphics*[width=72mm]{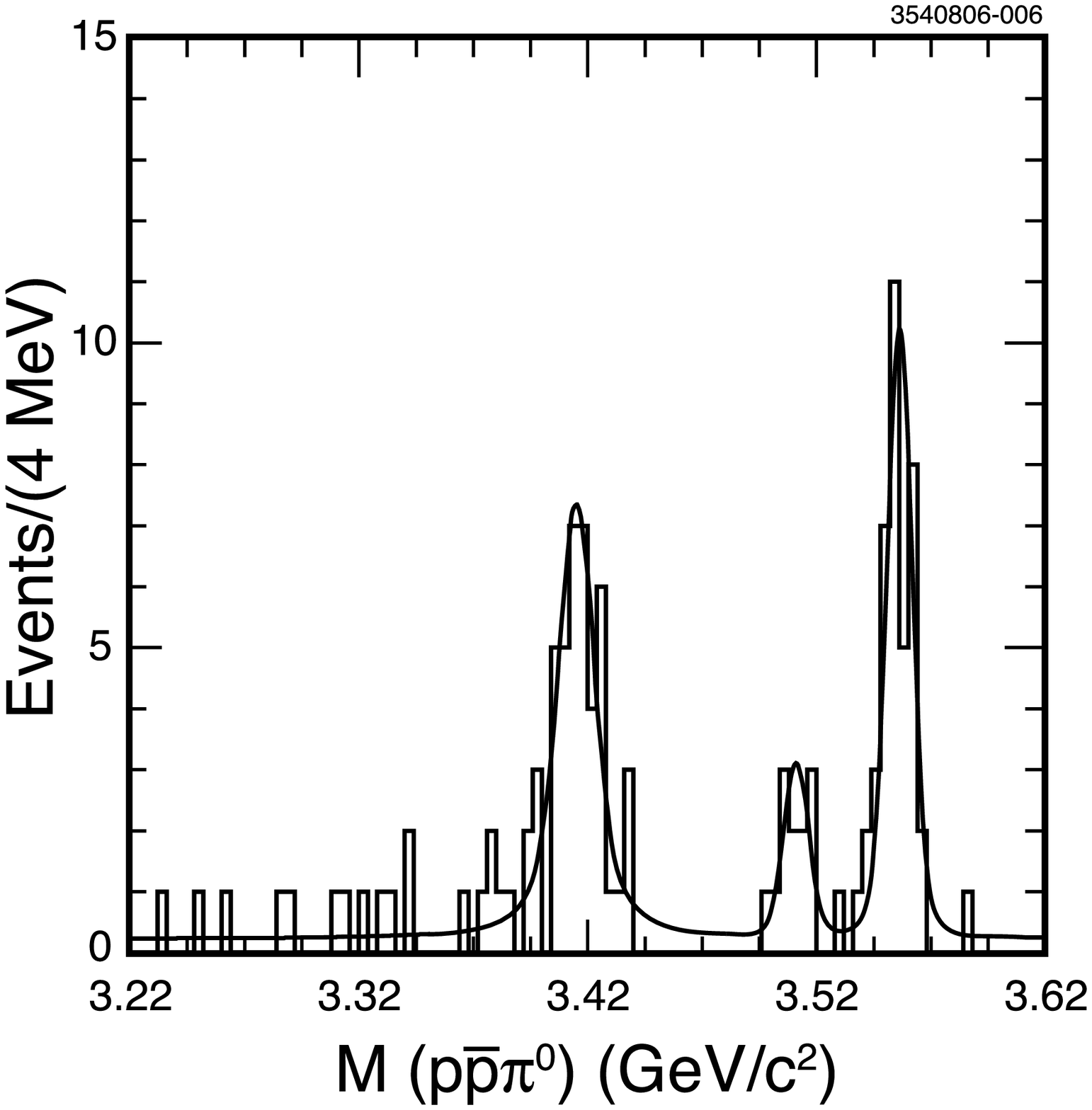}
\caption{Mass distribution for candidate $\chi_{cJ} \to p {\bar p} \pi^0$ events.
         The displayed fit is described in the text.}
\label{fig:ppbpi0}
\end{minipage}
\begin{minipage}[t]{75mm}
\includegraphics*[width=72mm]{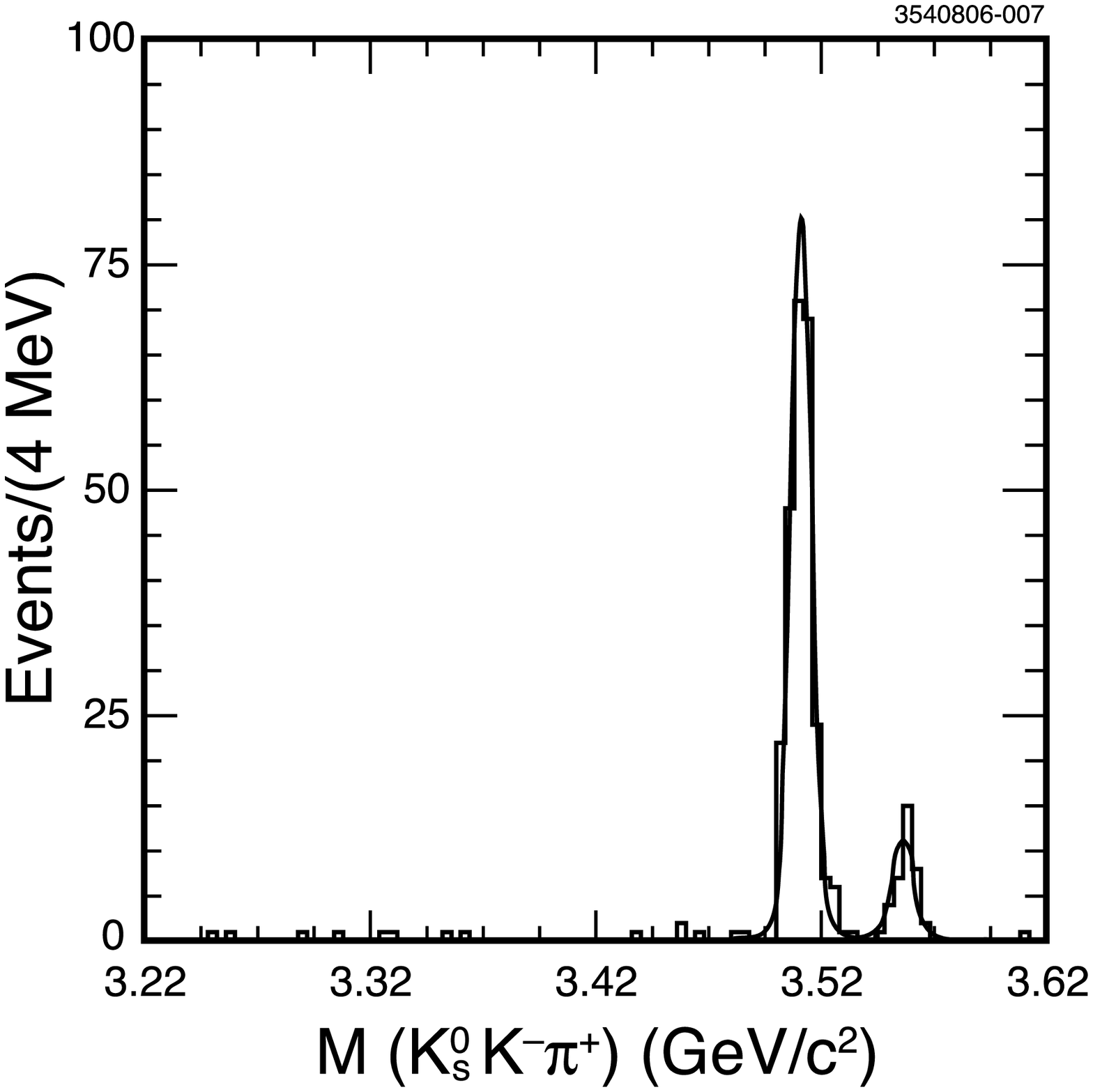}
\caption{Mass distribution for candidate $\chi_{cJ} \to K^0_{\rm S}K^-\pi^+$ events.
         The displayed fit is described in the text.}
\label{fig:piKKs}
\end{minipage}
\hfill
\begin{minipage}[t]{75mm}
\includegraphics*[width=72mm]{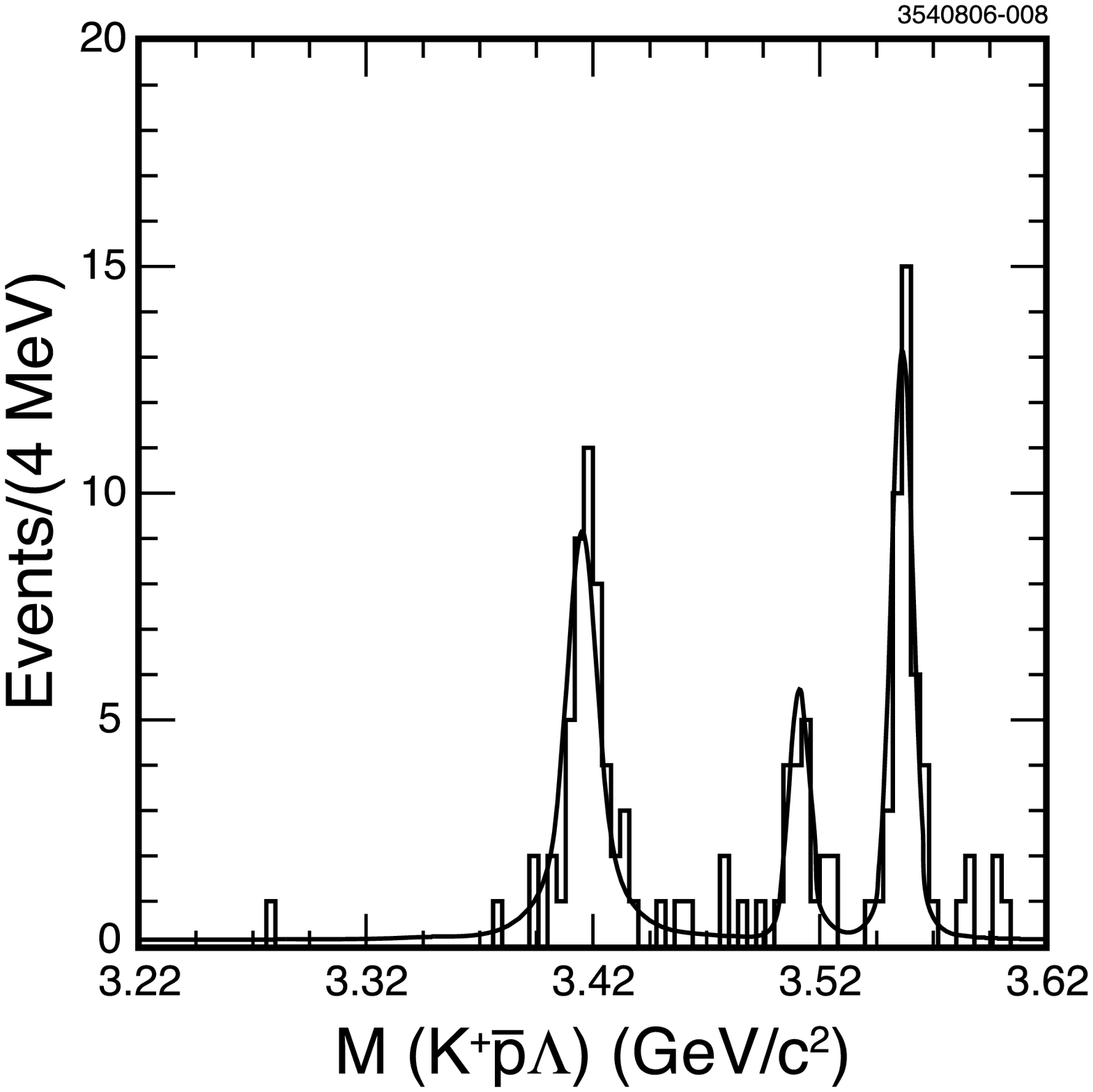}
\caption{Mass distribution for candidate $\chi_{cJ} \to K^+{\bar p}\Lambda$ events.
         The displayed fit is described in the text.}
\label{fig:Kpblam}
\end{minipage}
\end{figure}
Figures \ref{fig:pipieta}-\ref{fig:Kpblam} show the mass
distributions for the eight $\chi_{cJ}$ decay modes selected by
the analysis described above.  Signals are evident in all
three $\chi_{cJ}$ states, but not in all the modes.  Backgrounds
are small.  The mass distributions are fit to three signal shapes,
Breit-Wigners convolved with Gaussian detector resolutions, and a linear
background.  The $\chi_{cJ}$ masses and intrinsic 
widths are fixed at the values from the Particle
Data Group compilation~\cite{pdg}.  The detector resolution is taken
from the simulation discussed above.  The simulation
properly takes into account the amount of data in 
the two detector configurations, and the distribution
of different decay modes we have observed.  We approximate
the resolution with a single Gaussian distribution, and variations
are considered in the determination of systematic uncertainty.
The detector resolution dominates for the $\chi_{c1}$ and $\chi_{c2}$,
but is similar to the intrinsic width of the $\chi_{c0}$.
The fits are displayed in Figures \ref{fig:pipieta}-\ref{fig:Kpblam}
and summarized in Tables~\ref{tab:BRfits0}-\ref{tab:BRfits2}.
\begin{table}
\caption{Parameters used and results of fits to the $\chi_{cJ}$
         mass distributions of Figures~\ref{fig:pipieta}-\ref{fig:Kpblam} for the $\chi_{c0}$
         signal. The
         fit is described in the text and the yield error is only statistical.
         If no significant signal is observed we also show an upper limit at 90\% confidence level.}
\begin{tabular}{l|c|c|l}
\hline \hline
Mode                     & Efficiency (\%) & Resolution (MeV) & Yield \\ \hline
$\pi^+\pi^-\eta$         & 18.2            & 6.23             & $4.2\pm4.1$ ($<10.6$) \\
$K^+K^-\eta$             & 13.8            & 6.10             & $3.1\pm2.7$ ($<9.3$) \\
$p{\bar p}\eta$          & 16.1            & 5.42             & $17.7^{+5.2}_{-4.8}$ \\
$\pi^+\pi^-\eta^\prime$  &  8.0            & 4.38             & $2.0\pm3.8$ ($<8.5$) \\
$K^+K^-\pi^0$            & 27.6            & 6.47             & $-3.3\pm2.7$ ($<4.7$)\\
$p{\bar p}\pi^0$         & 27.8            & 5.80             & $46.4^{+8.0}_{-7.2}$ \\
$\pi^+K^-K^0_{\rm S}$    & 19.8            & 4.75             & $0.0\pm1.0$ ($<2.7$) \\
$K^+{\bar p}\Lambda$     & 16.8            & 4.38             & $51.3^{+8.1}_{-7.4}$ \\
\hline \hline
\end{tabular}
\label{tab:BRfits0} 
\end{table}
\begin{table}
\caption{Parameters used and results of fits to the $\chi_{cJ}$
         mass distributions of Figures~\ref{fig:pipieta}-\ref{fig:Kpblam} for the $\chi_{c1}$
         signal. The
         fit is described in the text and the yield error is only statistical.
         If no significant signal is observed we also show an upper limit at 90\% confidence level.}
\begin{tabular}{l|c|c|l}
\hline \hline
Mode                     & Efficiency (\%) & Resolution (MeV) & Yield \\ \hline
$\pi^+\pi^-\eta$         & 18.8            & 5.85             & $255^{+17}_{-16}$    \\
$K^+K^-\eta$             & 14.6            & 5.66             & $14.1^{+4.6}_{-3.9}$ \\
$p{\bar p}\eta$          & 17.7            & 5.41             & $3.2\pm2.3$ ($<7.6$) \\
$\pi^+\pi^-\eta^\prime$  &  8.5            & 4.37             & $57.6^{+8.4}_{-7.7}$ \\
$K^+K^-\pi^0$            & 29.2            & 6.79             & $157\pm13$           \\
$p{\bar p}\pi^0$         & 30.1            & 5.23             & $9.9^{+3.8}_{-3.2}$  \\
$\pi^+K^-K^0_{\rm S}$    & 20.6            & 4.37             & $249\pm16$           \\
$K^+{\bar p}\Lambda$     & 17.7            & 4.38             & $16.3^{+4.7}_{-4.0}$ \\
\hline \hline
\end{tabular}
\label{tab:BRfits1}
\end{table}
\begin{table}
\caption{Parameters used and results of fits to the $\chi_{cJ}$
         mass distributions of Figures~\ref{fig:pipieta}-\ref{fig:Kpblam} for the $\chi_{c2}$
         signal. The
         fit is described in the text and the yield error is only statistical.
         If no significant signal is observed we also show an upper limit at 90\% confidence level.}
\begin{tabular}{l|c|c|l}
\hline \hline
Mode                     & Efficiency (\%) & Resolution (MeV) & Yield \\ \hline
$\pi^+\pi^-\eta$         & 18.5            & 5.58             & $26.2^{+6.4}_{-5.7}$  \\
$K^+K^-\eta$             & 14.6            & 5.53             & $6.9\pm2.9$ ($<12.5$)\\
$p{\bar p}\eta$          & 17.2            & 5.08             & $9.5^{+3.8}_{-3.0}$   \\
$\pi^+\pi^-\eta^\prime$  &  8.5            & 4.32             & $12.4^{+4.8}_{-4.1}$  \\
$K^+K^-\pi^0$            & 27.5            & 6.85             & $24.8^{+5.8}_{-5.1}$  \\
$p{\bar p}\pi^0$         & 29.3            & 5.10             & $37.1^{+6.7}_{-6.1}$  \\
$\pi^+K^-K^0_{\rm S}$    & 20.2            & 4.45             & $36.8^{+6.6}_{-5.9}$  \\
$K^+{\bar p}\Lambda$     & 17.5            & 4.32             & $42.1^{+7.2}_{-6.5}$  \\
\hline \hline
\end{tabular}
\label{tab:BRfits2}
\end{table}
Note that for the $\chi_{c0}$ in Table~\ref{tab:BRfits0} the five modes for which no
significant signal is found (yield less than three standard deviations from zero)
are forbidden by parity conservation.

\begin{figure}
%\begin{tabular}{ccc}
%\includegraphics*[width=52mm]{etapipifit.eps} &
%\includegraphics*[width=52mm]{pi0KKfit.eps} &
%\includegraphics*[width=52mm]{kshkpifit.eps}
%\end{tabular}
\includegraphics*[width=160mm]{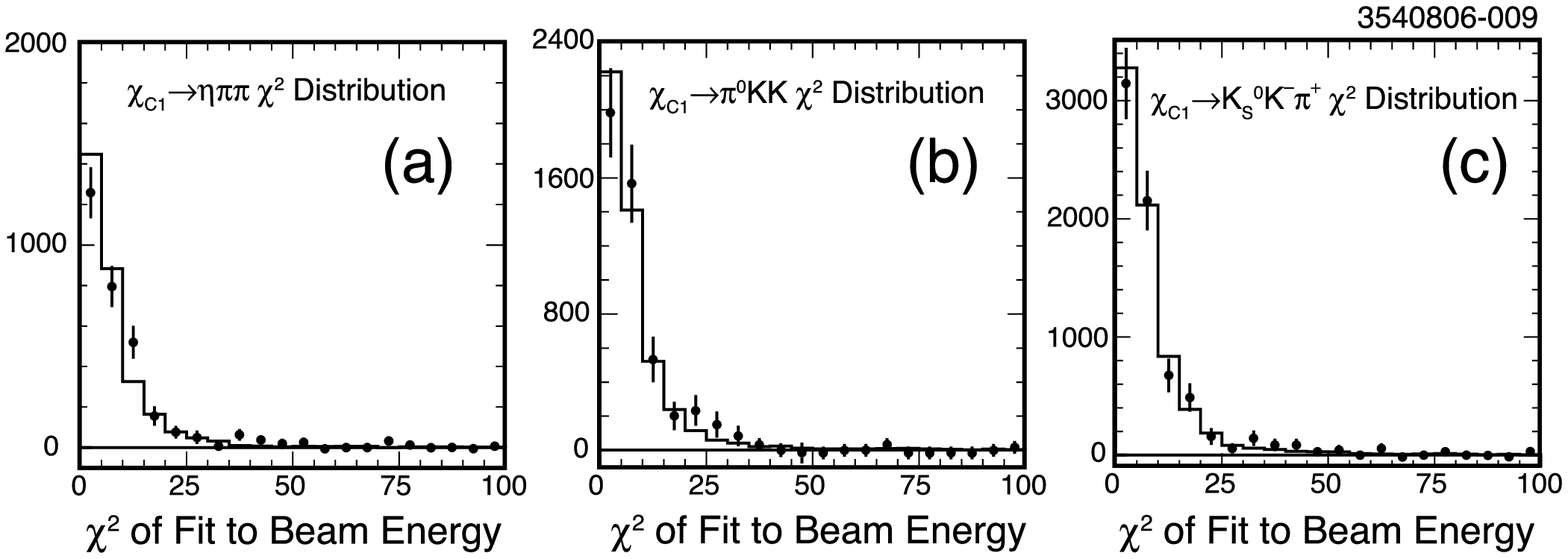}
\caption{Distribution of the $\chi^2$ for the event 4-momentum kinematic fit
         shown here with a $\chi_{c1}$ mass cut and sideband subtraction.  
         Plot (a) is the $\pi^+\pi^-\eta$ mode, (b) is the $K^+K^-\pi^0$ mode,
         and (c) is the $\pi^+K^-K^0_{\rm S}$ mode.  
         The data are shown by points and the simulation of signal events is
                        shown by the solid line normalized to have the same area.}
\label{fig:chi2}
\end{figure}

	We consider various sources of systematic uncertainties on the yields.
We varied the fitting procedure by allowing the $\chi_{cJ}$ masses and intrinsic widths
to float.  The fitted masses and widths agree with the values from the Particle
Data Group~\cite{pdg}, and we take the maximum variation in the observed yields, 
$\pm$4\%, as a systematic
uncertainty from the fit procedure.  Allowing a curvature term to the background
has a negligible effect.
For modes with large yields we can break up the sample into CLEO~III and CLEO-c
data sets, and fit with resolutions and efficiencies appropriate for the individual
data sets.  We note that the separate data sets give consistent 
efficiency-corrected 
yields and the summed yield differs by 2\% from the standard procedure,
which is small
compared to the $\pm$8\% statistical uncertainty.  We take this as the systematic
uncertainty from our resolution model.  From studies of other processes
we assign a $\pm$0.7\% uncertainty for the efficiency of finding each charged
track, $\pm$4.0\% for the $\gamma\gamma$ resonances, $\pm$2.0\% for each extra photon,
$\pm$1.3\% for the particle identification for each $K$ and $p$, $\pm$2.0\% for
secondary vertex finding, and $\pm$3.0\% from the statistical uncertainty
on the efficiency determined from the simulation.  We study the cut on the $\chi^2$
of the event 4-momentum kinematic fit in the three large yield $\chi_{c1}$ signals
by removing the $\chi^2$ cut, selecting events around the $\chi_{c1}$
mass peak, subtracting a low-mass side band, the only one available, and 
comparing the simulated $\chi^2$ distribution for signal events with the data
distribution.  This comparison is shown in Figure~\ref{fig:chi2}.  The agreement
between the data and simulation is good, and comparing the inefficiency introduced
by our cut on the 4-momentum kinematic fit $\chi^2$ between the data and the simulation
we assign a $\pm$3.5\% uncertainty on the efficiency due to uncertainty in modeling
this $\chi^2$ distribution.  
The simulation was generated assuming three-body phase-space for the $\chi_{cJ}$ decay products. 
Deviations from this are to be expected.  Based on the results of the Dalitz plot analyses
discussed below we correct the efficiency in the 
$\chi_{c1} \to \pi^+ \pi^-\eta $, 
$\chi_{c1} \to K^+K^-\pi^0$, and
$\chi_{c1} \to K^-\pi^+K^0_S$ 
modes by a relative $-2.2$\%, $-1.5$\%, and $+6$\% respectively 
to account for the change
in the efficiency caused by the deviation from a uniform phase space distribution
of decay products to what we actually observe.  
%This correction has
%a noticeable impact only on the $\chi_{c1} \to K^-\pi^+K^0_S 
We apply an additional
$\pm5$\% uncertainty on all other modes to account for the effect of resonant sub-structure
on the efficiency.  The total systematic uncertainty is mode dependent, but is roughly
10\%; it is higher for modes with many photons in the final state and lower for those with
only one.

To calculate $\chi_{cJ}$ branching fractions, we use previous CLEO measurements
for the $\psi(2S)\to\gamma\chi_{cJ}$
branching fractions from Equations~\ref{eqn:br_chic0}-\ref{eqn:br_chic2}
\cite{chicbf}.  
The uncertainties on these branching fractions
are included in the systematic uncertainty on the $\chi_{cJ}$ branching fractions we report.
Also there is a 3\% uncertainty on the number of $\psi(2S)$ produced.
Results for the three-body branching fractions are shown in 
Table~\ref{tab:Branching_fractions}.  Where the yields do not show clear signals
\begin{table}
\caption{\label{tab:Branching_fractions}
Branching fractions in % \%
units of 10$^{-3}$.  Uncertainties are statistical,
systematic due to detector effects plus analysis methods,
and a separate systematic due to uncertainties in the $\psi(2S)$ branching
fractions.  Limits are at the 90\% confidence level.}
%\small
\begin{tabular}{l|c|c|c}
\hline \hline
Mode &$\chi_{c0}$ & $\chi_{c1}$ & $\chi_{c2}$ \\
\hline
$\pi^+\pi^-\eta $         & $<0.21$  
                          & $5.0 \pm 0.3 \pm 0.4 \pm 0.3$ 
                          & $0.49\pm 0.12\pm 0.05\pm 0.03$ \\
$K^+K^-\eta$              & $<0.24$  
                          & $0.34\pm 0.10\pm 0.03\pm 0.02$ 
                          & $<0.33$                        \\
$p\bar{p}\eta$            & $0.39\pm 0.11\pm 0.04\pm 0.02$  
                          & $<0.16$ 
                          & $0.19\pm 0.07\pm 0.02\pm 0.01$ \\
$\pi^+\pi^-\eta^{\prime}$ & $<0.38$  
                          & $2.4 \pm 0.4  \pm0.2 \pm 0.2$     
                          & $0.51\pm 0.18 \pm0.05\pm 0.03$ \\
$K^+K^-\pi^0$             & $<0.06$  
                          & $1.95\pm 0.16\pm 0.18\pm 0.14$ 
                          & $0.31\pm 0.07\pm 0.03\pm 0.02$ \\
$p\bar{p}\pi^0$           & $0.59\pm 0.10\pm 0.07\pm 0.03$ 
                          & $0.12\pm 0.05\pm 0.01\pm 0.01$ 
                          & $0.44\pm 0.08\pm 0.04\pm 0.03$ \\
$\pi^+K^-\overline{K}^0$  & $<0.10$  
                          & $8.1 \pm 0.6 \pm 0.6 \pm 0.5$ 
                          & $1.3 \pm 0.2 \pm 0.1 \pm 0.1$  \\
$K^+ \bar{p}\Lambda$      & $1.07\pm 0.17\pm 0.10\pm 0.06$ 
                          & $0.33\pm 0.09\pm 0.03\pm 0.02$ 
                          & $0.85\pm 0.14\pm 0.08\pm 0.06$ \\ 
\hline \hline
\end{tabular}
\end{table}
we calculate 90\% confidence level upper limits using the yield 
central values with the statistical errors from the yield fits
combined in quadrature with the systematic uncertainties on the 
efficiencies and other branching fractions.  We assume
the uncertainty is distributed as a Gaussian and the upper limit 
is the branching fraction value at which 90\% of the
integrated area of the Gaussian falls below.  We exclude the 
unphysical region, negative branching fractions, for this
upper limit calculation.  We note that the ratio of rates obtained 
from isospin symmetry
(see Appendix~\ref{sec:Clebsch-Gordan},
Equations~\ref{eqn:case_width} and~\ref{eqn:case_width_for_a0}),
expected to be 4.0, is
consistent with our measurement:
\begin{equation} \label{eqn:ratio_of_rates}
  \frac{\Gamma(\chi_{c1}\to \pi^+K^-K^0)
       +\Gamma(\chi_{c1}\to \pi^-K^+\overline{K}^0) }
       {\Gamma(\chi_{c1}\to K^+K^-\pi^0)}
       = 4.2 \pm 0.7.
\end{equation}
Our results are consistent with branching fractions and upper limits
in the $\pi^+K^-K^0_{\rm S}$ and $\pi^+\pi^-\eta$ modes
from BES~\cite{BES}, but more precise.

%%%======================================================================
\section{Substructure analysis}

We perform a Dalitz plot analysis on the modes with the highest statistics,
$\chi_{c1} \to \pi^+ \pi^-\eta $, $\chi_{c1} \to K^+K^-\pi^0$, and
$\chi_{c1} \to \pi^+K^-K^0_S$, to study the two-body substructure.
For the Dalitz analysis only those
events within 10 MeV, roughly two standard deviations, of the 
observed $\chi_{c1}$ signal peak mean in the specific mode
are accepted.  For $\chi_{c1} \to \pi^+ \pi^-\eta$ there are 228 events in
this region and the signal fit finds 224.2 signal events and 5.1 
combinatorial background. For $\chi_{c1} \to K^+K^-\pi^0 $ there are
137 events accepted with the fit finding 137.8 signal and 2.4 background
events, and for $\chi_{c1} \to \pi^+K^-K^0_S$, the numbers
are 234 events, of which 233.2 are signal and 0.8 are background. In all
cases the contribution from the tail of the $\chi_{c2}$ is less than 
one event.

An unbinned maximum likelihood fit is used 
in order to perform the Dalitz plot analysis \cite{cleodalitz}. 
In order to assess the fit quality we use an adaptive binning technique 
\cite{D0-K0SETAPI0} and calculate a probability for Pearson statistics.
Efficiencies are determined with simulated event samples 
for the $\chi_{c1}$ decay generated uniformly in phase space,
and run through the analysis procedure described above.
The efficiency across the Dalitz plots is fit to a two-dimensional
polynomial of third order in the Dalitz plot variables.  The fits are
of good-quality and the efficiency is generally flat across the 
Dalitz plot.  

When fitting the data small contributions from backgrounds are neglected. 
We are examining the $e^+e^- \to \psi(2S) \to \gamma \chi_{c1}$ process.
In such a decay the $\chi_{c1}$ should be polarized.  In principle
a complete analysis would take into account the angle of the photon
with respect to the $e^+e^-$ beams' collision axis
and decompose the $\chi_{c1}$ decay into
its partial waves.  We use a simple model to 
analyze our data sample which is adequate for seeing the largest
contributions to the substructure in our small sample.
We take the Dalitz plot matrix element $|{\mathcal M}|^2$
to be a sum of non-interfering resonances,
\begin{equation}
  \label{eqn:sumofresonances}
        |{\mathcal M}|^2 = \sum_R |A_R|^2 \cdot \Omega_R^2 .
\end{equation}
The quantity $A_R(m)$ represents the amplitude
of each resonance contribution with angular distributions $\Omega_R^2$
taken from Ref.~\cite{Filippini-Fontana-Rotondi} as shown in 
Table~\ref{tab:angular_distributions}.
%
%%%================================================================
\begin{table}[!htb]
\caption{\label{tab:angular_distributions} Angular distributions from 
         Ref.~\cite{Filippini-Fontana-Rotondi} used in the present analysis. 
         The notation follows the original publication.
         We assume the decay of a particle with spin $J$ 
         ($J=1$ for $\chi_{c1}$ in our case)
         to the resonance $R$ of spin $j$ and a pseudoscalar particle 
         with relative orbital momentum $L$. 
         The angle $\theta$ is the resonance decay angle with respect to
         the pseudoscalar particle in the resonance's rest frame.
         The term $z^2=\gamma_R^2-1$ is a relativistic correction factor, where  
         $\gamma_R$ is the resonance Lorentz factor in the $\chi_{c1}$ rest frame.}
%\scriptsize
\begin{center}
\begin{tabular}{c|c}
\hline \hline
$J \to j+L$ & Angular distribution, $\Omega^2_R$ \\
\hline
1$\to$0+1 & uniform \\
1$\to$1+0 & $1 + z^2\cos^2\theta$ \\
%1$\to$1+1 & $\sin^2\theta$ \\
1$\to$1+2 & $1 + (3 + 4 z^2)\cos^2\theta$ \\
1$\to$2+1 & $(1+z^2) \big[1 + 3\cos^2\theta + 9 z^2 (\cos^2\theta - 1/3)^2 \big]$ \\
%1$\to$2+2 & $(1 + z^2)\cos^2\theta \sin^2\theta$     \\
\hline \hline
\end{tabular}
\end{center}
\end{table}
%%%================================================================
%
\\ Narrow resonances are described with a Breit-Wigner amplitude
\begin{equation}
  \label{eqn:BW}
 A_R(m) = \frac{a_R}{m_R^2 - m^2 - im \Gamma_{R,~\mathrm{total}}(m)}.
\end{equation}
The $a_R$ coefficients are fit parameters giving the amplitudes of the resonance 
with spin $j$- and mass-dependent width
\begin{equation}
  \label{eqn:mGamma}
    m \Gamma_{R,~\mathrm{total}}(m) = \sum_f {\mathcal B}(R \to f)\cdot m_R\Gamma_{R}
                               \bigg(\frac{p}{p_R}\bigg)^{2j+1}
                                     \frac{m_R}{m},
\end{equation}
where the 
resonance mass $m_R$, width $\Gamma_{R}$, and branching ratio ${\mathcal B}(R \to f)$
into the final state $f$ are taken from previous experiments \cite{pdg};
$p$ and $p_R$ are the decay products' momenta in the resonance rest frame
and its value at $m=m_R$.
For the scalar resonance $a_0(980)$ we use a Flatt\'e parameterization
in the style of the Crystal Barrel Collaboration~\cite{CBarrel_a0_980}, 
\begin{equation}
  \label{eqn:Flatte}
 A_{a_0(980)}(m) = \frac{a_{a_0(980)}}{ m_R^2 - m^2 - i[
                                                g^2_{\eta\pi} \rho_{\eta\pi}(m)
                                              + g^2_{\bar{K}K} \rho_{\bar{K}K}(m)
                                             ]
                            },
\end{equation}
where $m_R$, $g_{\eta\pi}$, and $g_{\bar{K}K}$ are the resonance mass
and coupling constants, $m$ is the invariant mass of the resonance final state, and
$\rho_{ab}(m)$ is a phase space factor for the particular final state. 
We use the $a_0(980)$ line-shape parameters from Table~\ref{tab:scalar_parameters}.
Similar details of the $f_0(980)$ parameterization
are unimportant as it is used only in systematic studies.

For low-mass $\pi^+\pi^-$($\sigma$) and  $K\pi$($\kappa$) $S$-wave contributions
we choose a parameterization with a complex pole $m_R$~\cite{Oller_2005}, 
  \begin{equation}
  \label{eqn:Oller}
         A_R(m) = \frac{a_R}{m_R^2 - m^2},
  \end{equation}
with $m_\sigma = (470 - i 220)$ MeV and $m_\kappa = (710 - i 310)$ MeV,
which is adequate for our small sample.

\begin{figure}
\includegraphics*[width=150mm]{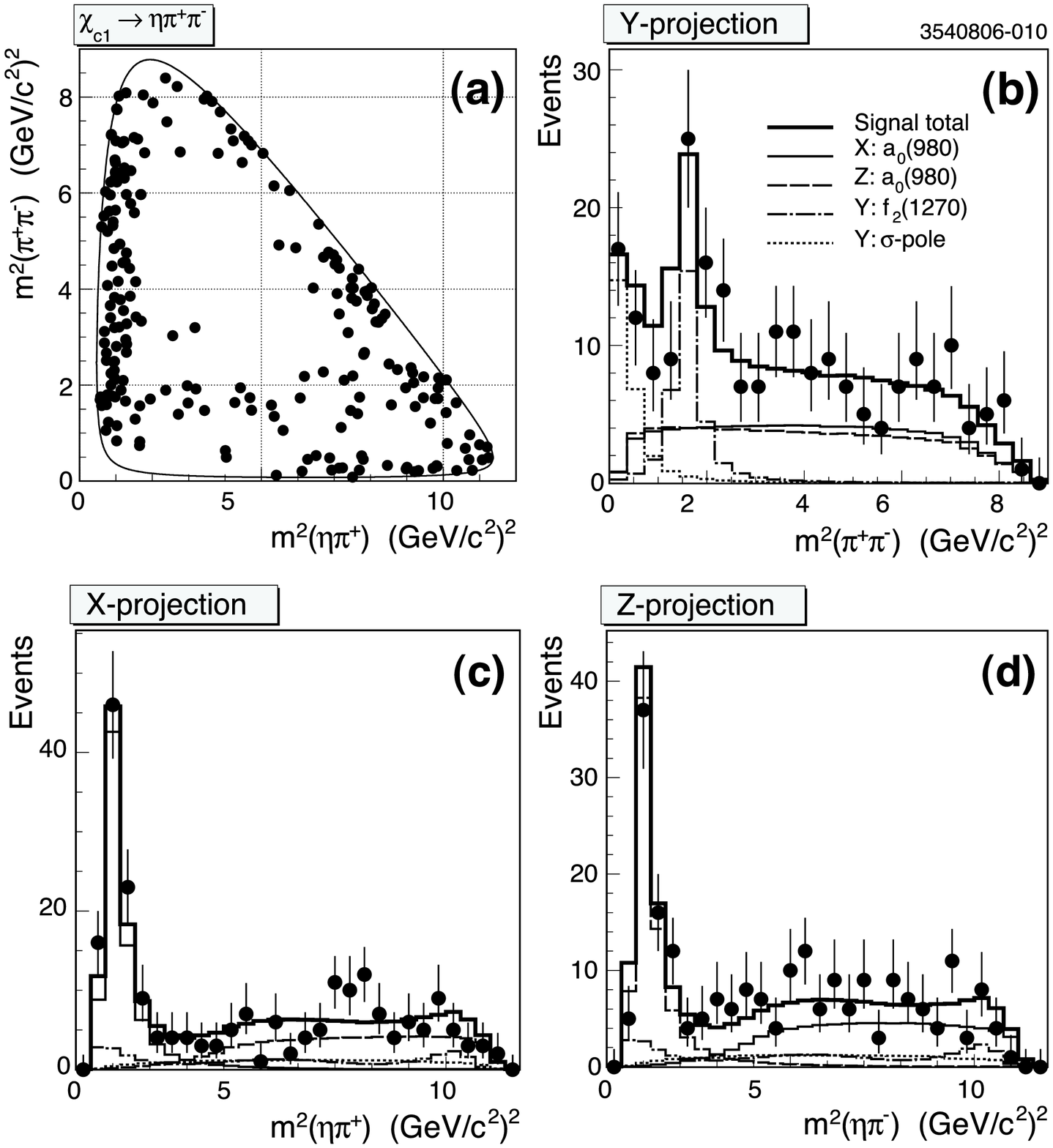}
\caption{(a) Dalitz plot, (b)--(d) projections on the three mass squared combinations
         for $\chi_{c1} \to \pi^+\pi^-\eta$.  The displayed fit projections
         are described in the text.}
\label{fig:pipietadalitz}
\end{figure}

	Figure~\ref{fig:pipietadalitz} shows the Dalitz plot
and three projections for $\chi_{c1} \to \pi^+\pi^-\eta$.
There are clear contributions from $a_0(980)^\pm \pi^\mp$ and $f_2(1270) \eta$ intermediate states,
and significant accumulation of events at low $\pi^+\pi^-$ mass.
Note that the $a_0(980)$ can contribute in two decay modes to the Dalitz plot.  
An isospin Clebsch-Gordan decomposition for this decay,
given in Appendix~\ref{sec:Clebsch-Gordan}, Equation~\ref{eqn:case_C_intermediate}, shows that 
amplitudes and strong phases of both charge-conjugated states should be equal.  
The overall amplitude normalization is
arbitrary and we set
$a_{a_0(980)^+} = a_{a_0(980)^-} = 1$.
All other fit components are defined relative
to this choice.

Our initial fit to this mode includes only $a_0(980)^\pm \pi^\mp$ and $f_2(1270) \eta$
contributions, but has a vanishing probability of 0.1\% for describing the data
due to the accumulation of events at low $\pi^+\pi^-$ mass.
To account for this we try $K^0_S$, $f_0(980)$, $\rho(770)$, and $\sigma$ resonances.
Only the $\rho(770)$ and the $\sigma$ give high fit probability. 
However the decay $\chi_{c1} \to \rho(770) \eta$ is $C$-forbidden,
and the low-mass distribution is not well represented by the $\rho(770)$, which
only gives an acceptable fit due to its large width and 
the limited statistics of our sample. 
The $\sigma$ describes well the low $\pi^+\pi^-$ mass spectrum,
and we describe the Dalitz plot with 
$a_0(980)^\pm \pi^\mp$, $f_2(1270) \eta$, and $\sigma \eta$ contributions.
Table~\ref{tab:pipietadalitz} gives the results of this fit, which has a 
probability to match the data of 66\%.
The angular distributions for $a_0(980)^\pm$ and $\sigma$ meson decays are uniform, 
and for $f_{2}(1270)$ are taken from 
Table~\ref{tab:angular_distributions} for quantum numbers $JjL=121$.
We assume that a possible contribution from $L=3$ is small and it is neglected.
\begin{table}[!htb]
\caption{\label{tab:scalar_parameters}  Resonance parameters
                            in the $\chi_{c1} \to \eta\pi^+\pi^-$ mode
                            comparing their nominal value to fit values
                            when individual resonance parameters
                            are allowed to float. }
%\scriptsize
\begin{center}
\begin{tabular}{c|c|c}
\hline \hline
Parameter                       & Nominal Value & When Floating \\
\hline
$m(a_0(980))$,  MeV/c$^2$       &  999          &  1002$\pm$18 \\
$g_{\eta\pi}$,  MeV/c$^2$       &  620          &  637$\pm$49  \\
$g_{K\bar{K}}$, MeV/c$^2$       &  500          &  523$\pm$154 \\
\hline
${\rm Re}(m_\sigma)$, MeV/c$^2$ &  470          &  511$\pm$28  \\
${\rm Im}(m_\sigma)$, MeV/c$^2$ &--220          &--102$\pm$50  \\
\hline \hline
\end{tabular}
\end{center}
\end{table}
\begin{table}
\caption{Fit results for $\chi_{c1} \to \eta\pi^+\pi^-$ Dalitz plot analysis.  The uncertainties
         are statistical and systematic.
         Allowing for interference among the resonances changes the fit fractions
         by as much as 20\% in absolute terms as discussed in the text.
        }
\begin{tabular}{l|c|c}
\hline \hline
Mode                   & $a_R$                    & Fit Fraction (\%)  \\ \hline
$a_0(980)^\pm \pi^\mp$ & 1                        & $75.1\pm3.5\pm4.3$ \\
$f_2(1270) \eta$       & $0.103\pm0.014\pm0.005$  & $14.4\pm3.1\pm1.9$ \\    
$\sigma \eta$          & $0.41\pm0.05\pm0.10$     & $10.5\pm2.4\pm1.2$ \\
\hline \hline
\end{tabular}
\label{tab:pipietadalitz} 
\end{table}
The systematic uncertainties shown in the table were obtained from
variations to this nominal fit as discussed below. 
We allow the 2D-efficiency to vary with its polynomial coefficients
constrained by the results of the fit to the simulated
events; the mass of the $a_0(980)$ and its coupling constants are allowed to float,
the parameters of the $\sigma$-pole are allowed to float, 
and we allow additional contributions
from $\rho(770) \eta$, $f_0(980) \eta$, $K^0_S \eta$, and $\pi_1(1400) \pi$.
The results of allowing the resonance parameters to float as compared
to their fixed values used in the nominal fit are shown in Table~\ref{tab:scalar_parameters}.
For the additional contributions we do
not observe amplitudes that are significant and we limit their 
individual fit fractions
to roughly less than 5\% of the Dalitz plot.
We note that with higher statistics this mode
may offer one of the best measurements of the parameters of the $a_0(980)$.
These results are consistent with the substructure analysis of this mode 
by BES~\cite{BES}.

\begin{figure}
\includegraphics*[width=150mm]{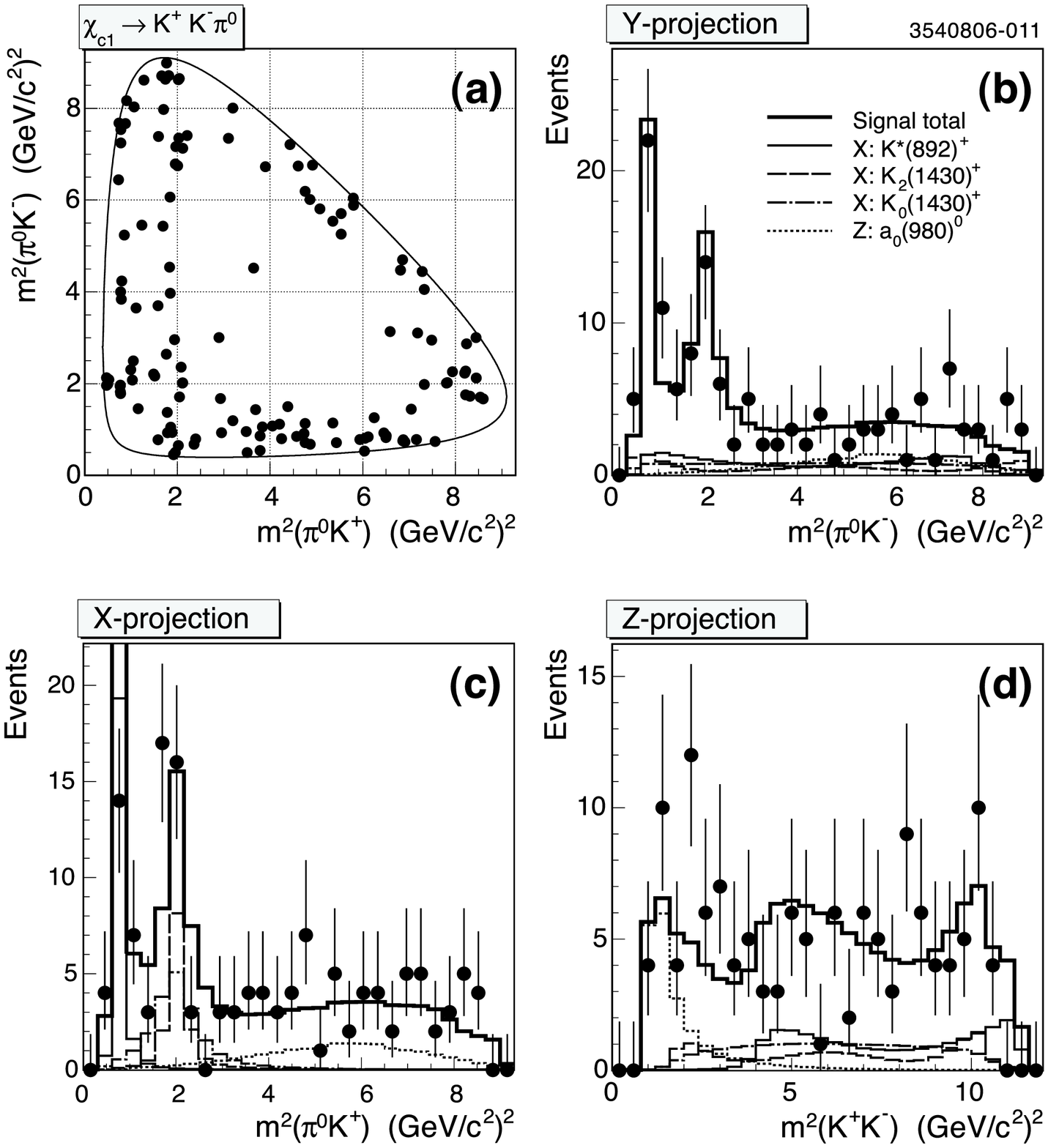}
\caption{(a) Dalitz plot, (b)--(d) projections on the three mass squared combinations
         for $\chi_{c1} \to K^+K^-\pi^0$.  The displayed fit projections
         are described in the text. 
         The contribution from $\pi^0K^-$ resonances are not shown. They look
         similar to the $\pi^0K^+$ resonance components for the relevant projections.}
\label{fig:KKpi0dalitz}
\end{figure}

\begin{figure}
\includegraphics*[width=150mm]{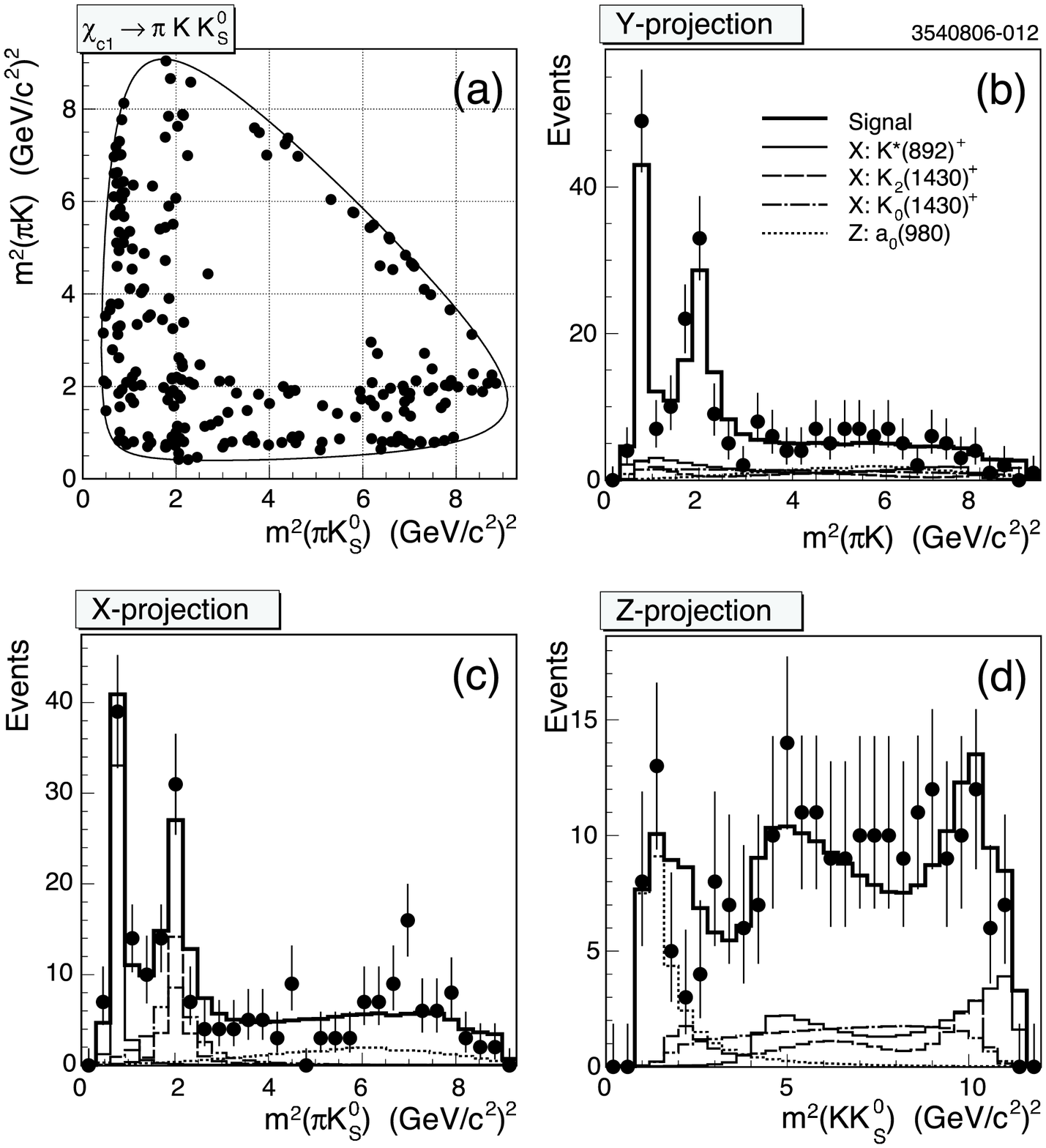}
\caption{(a) Dalitz plot, (b)--(d) projections on the three mass squared combinations
         for $\chi_{c1} \to \pi^+K^-K^0_S$.  The displayed fit projections
         are described in the text.
         The $\pi^+ K^-$ resonances are not shown. They look
         similar to the $\pi^+ K^0_S$  resonance components for the relevant projections.}
\label{fig:piKK0dalitz}
\end{figure}

The Dalitz plot for $\chi_{c1} \to K^+K^-\pi^0$ decay and
its projections are shown in Figure~\ref{fig:KKpi0dalitz}, and
for $\chi_{c1} \to \pi^+K^-K^0_S$ in Figure~\ref{fig:piKK0dalitz}.
We do a combined Dalitz plot analysis to these modes taking advantage
of isospin symmetry.  An isospin Clebsch-Gordan decomposition for these decays,
described in Appendix~\ref{sec:Clebsch-Gordan}, 
shows that these two Dalitz plots should have 
the same set of amplitudes for all  
$K^*\overline{K}$ and $a_0(980)\pi$ intermediate states. 
The relative factor $-\sqrt{2}$
between the two Dalitz plot amplitudes does not matter due to the 
individual normalization of their probability density functions.
In the combined fit to these two Dalitz plots, 
we use the following constraints on the amplitudes:
$a_{K^{*}} \equiv a_{K^{*+}} = a_{K^{*-}} = a_{K^{*0}} = a_{\overline{K}^{*0}}$,
and 
$a_{a(980)} \equiv a_{a(980)^+} = a_{a(980)^-} = a_{a(980)^0}$.
The overall amplitude normalization is arbitrary and
we set $a_{K^*(892)} = 1$.

Visual inspection shows apparent
contributions from $K^*(892)^\pm K^\mp$, $K^*(892)^0K^0_S$,
$K^*(1430)^\pm K^\mp$, $K^*(1430)^0K^0_S$, 
$a_0(980)^0\pi^0$, and $a_0(980)^\pm\pi^\mp$.
It is not clear if the $K^*(1430)$ are $K^*_0$ or $K^*_2$, and many
other $K\pi$ and $KK$ resonances can possibly contribute.  
The angular distribution for scalar resonance decays is always taken as uniform.
The shape for $K^*(892)K$ is taken from 
Table~\ref{tab:angular_distributions} for quantum numbers $JjL=110$.
We also modeled this contribution with the $JjL=112$ angular distribution
with the difference taken as a systematic uncertainty.
The shape for $K_2^*(1430)K$ is taken for quantum numbers $JjL=121$,
with a possible contribution from $L=3$ ignored.
Our best fit result is shown in Table~\ref{tab:KKpiDalitz}, displaying
\begin{table}[!htb]
\caption{Results of the combined fits to the 
$\chi_{c1} \to K^+ K^- \pi^0$ and $\chi_{c1} \to \pi K K^0_S$ Dalitz plots.
         Allowing for interference among the resonances changes the fit fractions
         by as much as 15\% in absolute terms as discussed in the text.}
\begin{tabular}{l|c|c}
\hline \hline
Mode                   & $a_R$                    & Fit Fraction (\%)   \\ \hline
$K^*(892)K$            & 1                        & $31.4\pm2.2\pm1.7$  \\
$K^*_0(1430)K$         & $3.8\pm0.4\pm0.2$        & $30.4\pm3.5\pm3.7$ \\
$K^*_2(1430)K$         & $0.44\pm0.06\pm0.04$     & $23.1\pm3.4\pm7.1$  \\
$a_0(980)\pi$          & $6.1\pm0.6\pm0.6$        & $15.1\pm2.7\pm1.5$  \\
\hline \hline
\end{tabular}
\label{tab:KKpiDalitz}
\end{table}
statistical and systematic errors.  
This fit has a good probability of matching the data, 73\%,
and agrees with fits done to the separate Dalitz plots not taking advantage of
isospin symmetry. The addition of a $\kappa K$ contribution
does not improve the fit probability and the amplitude $a_{\kappa}$
is statistically significant only at the four standard deviation level.
Similar behavior is noted for a non-resonant contribution.
Systematic uncertainties are evaluated by observing deviations from the
nominal fit as described above for the $\chi_{c1} \to \pi^+\pi^-\eta$ analysis.
When systematics are taken into account neither the $\kappa K$ nor
non-resonant contribution has a fit fraction significant at the three
standard deviation level.  Their individual contributions are limited to
less than roughly 10\% of the Dalitz plot.
These observations are consistent with those by
BES~\cite{BES} in the $\pi^+K^-K^0_{\rm S}$ mode.

While this model of the contributions to the Dalitz
plots gives a good description of the data it is clearly
incomplete.  There should be interference among the resonances,
which the simple approach of Equation~\ref{eqn:sumofresonances}
does not take into account. 
To quantify the effect of interference we have repeated the Dalitz
analyses using a ``quasi-coherent'' 
(note that $\Omega_R^2$ is always positively-defined)
sum of amplitudes with floating complex weighting factors instead of $a_R$.
This causes the
fit fractions to change by as much as $\sim$20\% absolutely 
in the $\eta\pi\pi$ analysis
and $\sim$15\% absolutely in the $K\overline{K}\pi$ analysis.
With more data a study of the $\chi_{c1}$ decay substructure
taking into account the effect of $\chi_{c1}$ polarization,
using the orientation of the $\chi_{c1}$ decay plane with respect to the $\chi_{c1}$
flight direction,
in a partial wave analysis would be a complete description of these decays.
The matrix element amplitudes should include the partial wave dependent
angular distributions and interference effects among the resonances.
See Ref.~\cite{BES_PWA_formalism} for appropriate prescriptions.
%
%%%%% DAVE'S initial VERSION:
%This addition to the model is still
%an incomplete description of the Dalitz plot as it does
%not take into account the effect of 
%as mentioned above.
%
%%%%% ANDER''S VERSION:
%In more detailed studies of the $\chi_{c1}$ Dalitz plots
%the matrix elements that incorporate interference
%between resonances need to be used.  In addition to looking
%at the distributions on the Dalitz plot, information could also
%be used about the orientation of the chi_c1 decay plane with
%respect to the chi_c1 flight direction.
%
%%%%% MIKHAIL's VERSION
%With higher statistics a more detailed study of the $\chi_{c1}$ decay substructure 
%in a partial wave analysis should be used in stead of the Dalitz plot analysis. 
%The $\chi_{c1}$ polarization has to be properly accounted. 
%The matrix element amplitude needs to incorparate spin dependent
%angular distribution and interference effects between resonances.
%Appropriate prescriptions can be found in \cite{BES_PWA_formalism}.

\section{Summary}

We have searched for and studied selected three-body hadronic decays
of the $\chi_{c0}$, $\chi_{c1}$, and $\chi_{c2}$ produced in radiative decays
of the $\psi(2S)$ in $e^+e^-$ collisions observed with the CLEO detector.  
Many of the channels covered in this analyses are observed or limited for
the first time.  Our observations and branching fraction limits are summarized in
Table~\ref{tab:Branching_fractions}.
In $\chi_{c1} \to \pi^+\pi^-\eta$ we have studied the resonant
substructure using
a Dalitz plot analysis simply modeling the resonance contributions
as non-interfering amplitudes.
Our results are summarized in Table~\ref{tab:pipietadalitz}.
We observe clear signals for the $a_0(980)^\pm \pi^\mp$ and $f_2(1270)\eta$
intermediate states,
and a low-mass $\pi^+\pi^-$ enhancement.
Similarly for $\chi_{c1} \to K\overline{K}\pi$ our results are summarized
in Table~\ref{tab:KKpiDalitz}, assuming, based on isospin, that
the $\pi K K^0_S$ and $K^+K^-\pi^0$ plots are identical.
We observe $K^*(892) K$, $K^*_2(1430)K$ and likely $a_0(980)\pi$ 
contributions in the $K \overline{K} \pi$ Dalitz plots.
Other conclusions about S-wave contributions are likely
model-dependent.

%%%======================================================================
\section{Acknowledgments}
% CURRENT acknowledgements go here...
% download from the CLEO website 
% http://www.lns.cornell.edu/restricted/CLEO/analysis/ac_help/ack.html
% This is the current version:
We gratefully acknowledge the effort of the CESR staff
in providing us with excellent luminosity and running conditions.
D.~Cronin-Hennessy and A.~Ryd thank the A.P.~Sloan Foundation.
This work was supported by the National Science Foundation,
the U.S. Department of Energy, and
the Natural Sciences and Engineering Research Council of Canada.

%%%================================================================

\section{Appendix: Clebsch-Gordan decomposition for $\chi_{c1}$}
\label{sec:Clebsch-Gordan}

In order to constrain amplitudes and phases in $\chi_{c1}$ decays
we use a Clebsch-Gordan decomposition
of the $\chi_{c1}$ (the state with $|I=0,I_Z=0\rangle$) for possible 
isospin subsystems:
\begin{equation} \label{eqn:Kpi}
    \chi_{c1} \to (K \pi)_{I=1/2} {\overline K},
\end{equation}
\begin{equation} \label{eqn:Kbarpi}
    \chi_{c1} \to ({\overline K} \pi)_{I=1/2} K,
\end{equation}
\begin{equation} \label{eqn:KKbar}
    \chi_{c1} \to (K {\overline K})_{I=1} \pi.
\end{equation}
Below we use the
Clebsch-Gordan decomposition rules, $|J,M\rangle = \sum_f c_f |m_1,m_2\rangle_f$
from Ref.~\cite{pdg}. 

%%%----------------------------------------------
\subsection{Clebsch-Gordan decomposition for $K^* \to K\pi$ decays}

We assume that $K^*$ mesons with I=1/2 form two isodoublets: 
($K^{*+}$, $K^{*0}$) and ($\overline{K}^{*0}$, $K^{*-}$) 
with ($I_Z=\frac{1}{2}$,$I_Z=-\frac{1}{2}$) respectively.
%The Clebsch-Gordan decomposition rules for isospin states 
%of product particles $\mathbf{ 1 \times \frac{1}{2} }$ are
The Clebsch-Gordan decomposition rules for $\mathbf{ 1 \times \frac{1}{2} }$ isospin
states are:

%$|\frac{1}{2}, \frac{1}{2}\rangle = \sqrt{\frac{2}{3}}|1,-\frac{1}{2}\rangle 
%                            - \sqrt{\frac{1}{3}}| 0,\frac{1}{2}\rangle$

%$|\frac{1}{2},-\frac{1}{2}\rangle = \sqrt{\frac{1}{3}}|0,-\frac{1}{2}\rangle
%                            - \sqrt{\frac{2}{3}}|-1,\frac{1}{2}\rangle$

\begin{equation} \label{eqn:Kstar+}
     K^{*+} %= |I=\frac{1}{2}, I_z=\frac{1}{2}\rangle 
            = \sqrt{\frac{2}{3}} \pi^+K^0 
            - \sqrt{\frac{1}{3}} \pi^0 K^+,
\end{equation}
\begin{equation} \label{eqn:Kstar0}
     K^{*0} %= |I=\frac{1}{2}, I_z=-\frac{1}{2}\rangle 
            = \sqrt{\frac{1}{3}} \pi^0K^0 
            - \sqrt{\frac{2}{3}} \pi^- K^+,
\end{equation}
\begin{equation} \label{eqn:Kstar0bar}
     \overline{K}^{*0}
            %= |I=\frac{1}{2}, I_z=\frac{1}{2}\rangle 
            = \sqrt{\frac{2}{3}} \pi^+K^- 
            - \sqrt{\frac{1}{3}} \pi^0 \overline{K}^0,
\end{equation}
\begin{equation} \label{eqn:Kstar-}
     K^{*-} %= |I=\frac{1}{2}, I_z=-\frac{1}{2}\rangle 
            = \sqrt{\frac{1}{3}} \pi^0K^- 
            - \sqrt{\frac{2}{3}} \pi^- \overline{K}^0.
\end{equation}

%%%================================================================
\subsection{Cases of $\chi_{c1} \to (K \pi)_{I=1/2} {\overline K}$ and  
                     $\chi_{c1} \to ({\overline K} \pi)_{I=1/2} K$  decays }
\label{ref:Kpi_I05}
For $\chi_{c1} \to K^*\overline{K}$ and 
    $\chi_{c1} \to \overline{K}^* K$ modes we use the 
$\mathbf{ \frac{1}{2} \times \frac{1}{2} }$ rule: 
% $|0,0\rangle=\frac{1}{\sqrt{2}}(|\frac{1}{2},-\frac{1}{2}\rangle - |-\frac{1}{2},\frac{1}{2}\rangle )$
\begin{equation} \label{eqn:case_A}
     \chi_{c1} = \frac{1}{\sqrt{2}}(K^{*+} K^- - K^{*0} \overline{K}^0),
\end{equation}
\begin{equation} \label{eqn:case_B}
     \chi_{c1} = \frac{1}{\sqrt{2}}(\overline{K}^{*0}K^0 - K^{*-} K^+).
\end{equation}

%%%----------------------------------------------

Combining Equations~\ref{eqn:case_A} and \ref{eqn:case_B} 
with Equations~\ref{eqn:Kstar+}-\ref{eqn:Kstar-} we get
\begin{equation} \label{eqn:case_A_FS}
     \chi_{c1}\sqrt{2} =
               \sqrt{\frac{2}{3}} \bigg[ (\pi^+K^0)K^- + (\pi^-K^+)\overline{K}^0 \bigg]
             - \sqrt{\frac{1}{3}} \bigg[ (\pi^0K^+)K^- + (\pi^0K^0)\overline{K}^0 \bigg],
\end{equation}
\begin{equation} \label{eqn:case_B_FS}
     \chi_{c1}\sqrt{2} = 
               \sqrt{\frac{2}{3}} \bigg[ (\pi^+K^-)K^0 + (\pi^-\overline{K}^0)K^+ \bigg]
             - \sqrt{\frac{1}{3}} \bigg[ (\pi^0K^-)K^+ + (\pi^0\overline{K}^0)K^0 \bigg].
\end{equation}
Assuming charge symmetry the amplitudes in 
Equations~\ref{eqn:case_A_FS} and \ref{eqn:case_B_FS} should be equal.
From these equations we get the ratio of rates:
\begin{equation} \label{eqn:case_rate_R1}
    \Gamma(\chi_{c1}\to \pi^+K^-K^0) /
    \Gamma(\chi_{c1}\to K^+K^-\pi^0) = 2, 
\end{equation}
\begin{equation} \label{eqn:case_rate_R2}
    \Gamma(\chi_{c1}\to \pi^-K^+\overline{K}^0) /
    \Gamma(\chi_{c1}\to K^+K^-\pi^0) = 2,
\end{equation}
or their sum
\begin{equation} \label{eqn:case_width}
             \Gamma(\chi_{c1}\to \pi^+K^-K^0)
           + \Gamma(\chi_{c1}\to \pi^-K^+\overline{K}^0) 
   = 4 \cdot \Gamma(\chi_{c1}\to K^+K^-\pi^0).
\end{equation}

%%%----------------------------------------------
\subsection{Case of $\chi_{c1} \to (K {\overline K})_{I=1} \pi$ decay}
\label{ref:KK_I1}

For $\chi_{c1} \to a\pi$ modes we use the $\mathbf{ 1\times 1 }$ rule: 
% $|0,0\rangle=\frac{1}{\sqrt{3}}(|1,-1\rangle - |0,0\rangle + |-1,1\rangle )$
\begin{equation} \label{eqn:case_C_intermediate}
     \chi_{c1} = \frac{1}{\sqrt{3}}(a^+\pi^- - a^0\pi^0 + a^-\pi^+)
\end{equation}
For $a\to K{\overline K}$ we use the $\mathbf{\frac{1}{2} \times \frac{1}{2}}$ rules:
% $a^+$: $|1,1\rangle=|\frac{1}{2},\frac{1}{2}\rangle$
\begin{equation} \label{eqn:case_a+}
         a^+ = K^+ \overline{K}^0,
\end{equation}
% $a^0$: $|1,0\rangle=\frac{1}{\sqrt{2}}(|\frac{1}{2},-\frac{1}{2}\rangle + |-\frac{1}{2},\frac{1}{2}\rangle )$
\begin{equation} \label{eqn:case_a0}
         a^0 = \frac{1}{\sqrt{2}} (K^+ K^- + K^0 \overline{K}^0),
\end{equation}
% $a^-$: $|1,-1\rangle=|-\frac{1}{2},-\frac{1}{2}\rangle$
\begin{equation} \label{eqn:case_a-}
         a^- = K^0 K^-.
\end{equation}
Combining Equation~\ref{eqn:case_C_intermediate} 
with Equations~\ref{eqn:case_a+}-\ref{eqn:case_a-} we get

\begin{equation} \label{eqn:case_C}
     \chi_{c1}\sqrt{3} = 
                      (K^+ \overline{K}^0)\pi^- 
                    - \frac{1}{\sqrt{2}} \bigg[(K^+ K^-)\pi^0  + (K^0 \overline{K}^0)\pi^0 \bigg]
                    + (K^0 K^-)\pi^+.
\end{equation}

From Equation~\ref{eqn:case_C} we get the ratio
of rates:
\begin{equation} \label{eqn:case_rate_R1_for_a0}
    \Gamma(\chi_{c1}\to \pi^+K^-K^0) /
    \Gamma(\chi_{c1}\to K^+K^-\pi^0) = 2, 
\end{equation}
\begin{equation} \label{eqn:case_rate_R2_for_a0}
    \Gamma(\chi_{c1}\to \pi^-K^+\overline{K}^0) /
    \Gamma(\chi_{c1}\to K^+K^-\pi^0) = 2,
\end{equation}
or their sum
\begin{equation} \label{eqn:case_width_for_a0}
             \Gamma(\chi_{c1}\to \pi^+K^-K^0)
           + \Gamma(\chi_{c1}\to \pi^-K^+\overline{K}^0) 
   = 4 \cdot \Gamma(\chi_{c1}\to K^+K^-\pi^0).
\end{equation}

%%%================================================================
\subsection{Consequences for Dalitz plot analysis}

Comparing 
Equations~\ref{eqn:case_rate_R1}-\ref{eqn:case_width}
for intermediate states with $K^*$ 
and
Equations~\ref{eqn:case_rate_R1_for_a0}-\ref{eqn:case_width_for_a0}
for intermediate states with $a(980)$,
we note that they are identical.
Thus observations of $\pi^+K^-K^0$ and 
charge conjugated $\pi^-K^+\overline{K}^0$ final states 
on the same Dalitz plot will yield the certain ratio
between $K^*$ and $a(980)$ amplitudes for the $\pi K K^0_S$ Dalitz plot.
The same ratio between $K^*$ and $a(980)$ amplitudes 
is expected for the $K^+K^-\pi^0$ Dalitz plot. 
This isospin analysis implies that these two Dalitz plots, 
$\pi K K^0_S$ and $K^+K^-\pi^0$,
can be parametrized using a common set of parameters for each 
$K^*$ and $a(980)$ intermediate state.
From Equations~\ref{eqn:case_A_FS}, \ref{eqn:case_B_FS} we can
write equations between decay amplitudes with $K^*$ mesons
\[
{\rm for~} \pi K K^0_S:~~~  
           a_{K^{*+}K^- \to K^0K\pi} 
         = a_{K^{*-}K^+ \to K^0K\pi} 
         = a_{K^{*0}\overline{K}^0 \to K^0K\pi}
         = a_{\overline{K}^{*0}K^0 \to K^0K\pi}
         =
\]
\begin{equation}
   \label{eqn:ampl_kstar}
\hspace{-20mm}
{\rm for~} K^+K^-\pi^0:~~~~~
         = -\frac{1}{\sqrt{2}} a_{K^{*+}K^- \to K^+K^-\pi^0}
         = -\frac{1}{\sqrt{2}} a_{K^{*-}K^+ \to K^+K^-\pi^0},
\end{equation}
where the signs assume equal phases
\begin{equation}
 \phi_{K^{*+}} = \phi_{K^{*-}} = \phi_{K^{*0}} = \phi_{\overline{K}^{*0}}.
\end{equation}
Similar equations between amplitudes with $a_0(980)$ can be obtained
from Equation~\ref{eqn:case_C}
\[
{\rm for~} \pi K K^0_S:~~~ 
           a_{a(980)^+\pi^- \to \pi K K^0} 
         = a_{a(980)^-\pi^+ \to \pi K K^0}
         =
\]
\begin{equation}
   \label{eqn:ampl_a0_980}
\hspace{-13mm}
{\rm for~} K^+K^-\pi^0:~~~
         = -\frac{1}{\sqrt{2}}a_{a(980)^0\pi^0 \to K^+K^-\pi^0},  
\end{equation}
assuming equal phases
\begin{equation}
 \phi_{a(980)^+} = \phi_{a(980)^-} = \phi_{a(980)^0}.
\end{equation}
Equations~\ref{eqn:ampl_kstar} and \ref{eqn:ampl_a0_980} predict that 
the ratio of amplitudes between these two Dalitz plots is $-\sqrt{2}$.
This relative factor does not matter, because each Dalitz plot
is normalized separately.
In the combined fit we ignore this $-\sqrt{2}$ factor 
%(along with branching fraction factors for $K^0 \to K^0_S \to \pi^+\pi^-$ transition) 
between the amplitudes 
in the $\pi K K^0_S$ and $K^+K^-\pi^0$ Dalitz plots, and use
common fit parameters for each of $K^{*}$ mesons and $a_0(980)$ intermediate states   
\begin{equation}
 a_{K^{*}} \equiv a_{K^{*+}} = a_{K^{*-}} = a_{K^{*0}} = a_{\overline{K}^{*0}},
\end{equation}
%\begin{equation}
% \phi_{K^{*}} \equiv \phi_{K^{*+}} = \phi_{K^{*-}} = \phi_{K^{*0}} = \phi_{\overline{K}^{*0}},
%\end{equation}
\begin{equation}
 a_{a(980)} \equiv a_{a(980)^+} = a_{a(980)^-} = a_{a(980)^0}.
\end{equation}
%\begin{equation}
% \phi_{a(980)} \equiv \phi_{a(980)^+} = \phi_{a(980)^-} = \phi_{a(980)^0}.
%\end{equation}

In Sections~\ref{ref:KK_I1} and \ref{ref:Kpi_I05} we have checked an isospin symmetry
between $\pi K K^0$ and $K^+K^-\pi^0$ decays
for two particular intermediate states. 
However, this symmetry is valid independently of
how the amplitudes are decomposed into two-body sub-amplitudes, and thus
should be valid for all points on the Dalitz plot.

%%%================================================================
%\section*{References}

\end{document}